\documentclass[journal]{IEEEtran}
\ifCLASSINFOpdf

\else

\fi

\usepackage{amsmath,amssymb,amsfonts}
\usepackage{graphicx}
\usepackage{textcomp}
\usepackage{xcolor}
\usepackage{cite}
\usepackage{multirow}
\usepackage{booktabs}
\usepackage{array}
\usepackage{url}

\usepackage{siunitx}
\usepackage{lineno}

\begin{document}

\title{{System-Level} Limits of Higher-Order QAM in Hollow-Core Fiber Systems}

\author{Md Ghulam Saber and Zhiping Jiang
\thanks{M. G. Saber and Z. Jiang are with the Ottawa Research Center, Huawei Technologies Canada, 303 Terry Fox Drive, Kanata, ON, K2K 3J1, Canada.}

\thanks{Manuscript received May 28, 2026; revised July 3, 2026 and July 20, 2026.}}

\markboth{Journal of Lightwave Technology}%
{Saber \MakeLowercase{\textit{et al.}}: {System-Level} Limits of Higher-Order QAM in Hollow-Core Fiber Systems}

\maketitle

\begin{abstract}
Hollow-core fiber (HCF) is widely expected to enable higher-order
quadrature amplitude modulation (QAM) through its near-vacuum Kerr
nonlinearity and higher launch power. We develop a per-channel
effective signal-to-noise ratio (SNR) budget that combines, reciprocally, the optical
link (amplified spontaneous emission (ASE), Kerr nonlinear interference (NLI), inter-modal interference (IMI), pigtail NLI, and
CO$_2$ gas absorption), a parameterized, symbol-rate-dependent
transceiver back-to-back (B2B) SNR ceiling (effective number of bits (ENOB) at rate, analog
bandwidth, Tx/Rx nonlinearity), and the remaining transceiver and
line impairments---laser phase noise, equalization-enhanced phase
noise, timing jitter, polarization-dependent loss, and amplifier
gain ripple with filter narrowing---each expressed as an equivalent
SNR floor. The central result, at a {representative} 64~GBaud
(75~GHz channel), is that once HCF removes the fiber limits the
transceiver ceiling---not the fiber---sets the achievable order:
a $\sim$25~dB ceiling at 64~GBaud makes 1024-QAM and above infeasible on
either fiber, confining ultra-high QAM to low baud. HCF's advantage is
therefore reach- and baud-at-a-given-order: at IMI coefficient, $\kappa=-55$~dB/km,
256-QAM reach grows from $\sim$45 to $\sim$170~km and 64-QAM from
$\sim$415 to $\sim$2275~km (single-mode fiber, SMF, $\rightarrow$ HCF). In the C-band the CO$_2$ lines are weak and sparse, so channels placed
off the lines follow the gas-free baseline and only worst-case
placements lose reach at long distances (256-QAM: $\sim$170 to
$\sim$90~km); in the L-band the strong absorption bands are denser than
the channel bandwidth, making line avoidance costly in spectrum, and a
channel-on-line loses one to two QAM orders. 
\end{abstract}


%
\IEEEpeerreviewmaketitle

\section{Introduction}
%
%
%
%

\IEEEPARstart{H}{ollow} core fiber (HCF) has emerged as a promising transmission medium for coherent optical systems because it combines very low Kerr nonlinearity, low chromatic dispersion, and reduced propagation delay relative to standard single-mode fiber (SMF)~\cite{Poletti2014,Jasion2022,saber2026_hcfimdd,Petrovich2025,saber2026ecoc,Fokoua2023}. These properties make HCF especially attractive for higher-order quadrature amplitude modulation (QAM), where achievable reach is dictated by the usable signal-to-noise ratio (SNR) remaining after fiber, line-system, and transceiver impairments are all accounted for. The practical appeal of HCF is therefore not merely lower nonlinearity by itself, but whether that physical advantage can be converted into higher spectral efficiency and longer reach at a given baud rate.

That promise, however, is not automatic. Once the Kerr limit is relaxed, other impairments can become dominant~\cite{saber2026_perspective}. In practical HCF links, the most important are inter-modal interference (IMI), discrete splice-related (both SMF-HCF and HCF-HCF) multipath interference (MPI), and the residual SMF pigtails inserted at amplifier and line-system interfaces \cite{Hahn2026_LPM,Hahn2026_OFC}. Transceiver penalties further constrain the usable modulation order, including laser phase noise, carrier-phase estimation (CPE) limits, equalization-enhanced phase noise (EEPN), finite effective number of bits (ENOB), timing jitter, and polarization-dependent loss (PDL)~\cite{Pfau2009,Savory2008,Sasai2020,Mazur2019,Shi2024_pilot,Shieh2008,Xie2009,Walden1999}. The question this paper addresses follows: as HCF demonstrations continue to push QAM order upward, do HCF-specific impairments and transponder-level implementation constraints set a practical ceiling well before the optical signal-to-noise ratio (OSNR) argument predicts, and which of those constraints binds first?

Existing studies do not yet give a complete answer to that question. Prior work has established the role of Kerr nonlinear interference (NLI) in SMF systems, characterized the key physical properties of HCF, and begun to quantify HCF-specific modal-coupling effects~\cite{Poggiolini2012,Poggiolini2015_eGN,Poggiolini2022_HCF,Fokoua2023,Nespola2021,saber2026protectionswitchinghybridhollowcore}. Other studies have examined individual transceiver penalties in coherent optical systems---primarily in conventional-fiber system contexts---including phase-noise tolerance, pilot-aided CPE, and quantization limits~\cite{Pfau2009,Sasai2020,Mazur2019,Shi2024_pilot,Walden1999,Varughese2018}. However, these effects are often treated in isolation, which obscures their relative importance on a common basis and makes it difficult to identify the true bottleneck for high-order QAM on HCF. What is still missing is a unified per-channel SNR budget that places fiber impairments, deployment penalties, and transceiver limitations on the same footing and translates them directly into feasible QAM order and reach.

The contribution of this paper is, to the best of our knowledge, the
first per-channel SNR budget that places all of the following on
a single footing for coherent C-band HCF high-order QAM, and combines
them in the physically correct parallel-noise form: (i)~the HCF-specific
optical-link impairments---fiber Kerr NLI, distributed and discrete IMI
with a differential modal attenuation (DMA)-aware all-pairs round-trip
MPI model, two-sided SMF-pigtail (and mode-field-adapter) NLI, gas-line
CO$_2$ absorption, and amplifier ripple/reconfigurable optical add-drop multiplexer (ROADM) line-system penalties; and
(ii)~a measurable, symbol-rate-dependent transceiver back-to-back SNR
ceiling that aggregates ENOB at the operating sample rate, finite analog
bandwidth, transmitter/receiver nonlinearity, EEPN (with digital
subcarrier multiplexing), order-dependent probabilistic-shaping gain,
bandwidth-weighted timing jitter, and PDL. The novelty is not any single
term but this particular combination, evaluated as a parallel
combination of a link SNR and a transceiver ceiling, which lets the
limiting constraint be identified directly in terms of feasible QAM
order and reach. The capacity-level study of Klaus and
Winzer~\cite{Klaus2022_HCF} treats receiver noise as a single lumped
floor; here we resolve that floor into measurable components and
combine it with the HCF-specific fiber terms.

The main picture that emerges differs from the conventional expectation.
HCF's near-vacuum Kerr nonlinearity and low dispersion indeed remove the
fiber limits, but once they are removed the binding constraint at
deployable symbol rates is the transceiver: a real coherent
transceiver delivers only a finite back-to-back SNR, which falls as the
symbol rate rises, and at a {representative} 64~GBaud this
back-to-back limit caps the achievable QAM order at 256-QAM-or-below
regardless of fiber type---a limit no fiber improvement can lift.
What HCF does deliver is longer
reach at a given order and a higher usable baud at a given order; among
the fiber terms the distributed IMI coefficient is the most
influential, with discrete splice MPI and practical SMF pigtails much
smaller in the operating regime of interest. Our aim is to put a
system-design lens on a set of impairments usually treated in isolation
in the HCF literature, and to identify where the true bottleneck lies.

Published experiments are increasingly consistent with this interpretation. On SMF, very high QAM orders {(1024- through 16384-QAM)} have been demonstrated {at low baud} in carefully optimized laboratory {settings~}\cite{Koizumi2012_1024QAM,Beppu2015_2048QAM,Olsson2018_4096QAM200km,Chen2019_16384QAM}. On HCF, {demonstrations} {have} established polarization-division-multiplexed (PDM) probabilistically-shaped (PS) 256-QAM (PDM-PS-256-QAM) over NANF~\cite{Wang2024_NANF} {and} PDM-PS-1024-QAM over a 5-km double-nested antiresonant nodeless fiber (DNANF) link at 20~GBaud~\cite{Fan2026_1024QAM_DNANF}{,} {while} {others} have emphasized {long-reach,} high-capacity transmission at more moderate modulation orders~\cite{Ge2025_AR_HCF,Hong2024_HCF_terabit}.

The remainder of this paper is organized as follows. Section~\ref{sec:system_model} describes the system architecture and parameters. Section~\ref{sec:impairment_models} presents the impairment models. Section~\ref{sec:results} presents numerical results. Section~\ref{sec:discussion} discusses implications and limitations, and Section~\ref{sec:conclusion} concludes the paper.

\section{System Model and Parameters}
\label{sec:system_model}

We consider a wavelength-division multiplexed (WDM) coherent system
with 80 channels at 75~GHz spacing, 64~GBaud per channel
(6~THz occupied bandwidth), dual polarization, and inline
erbium-doped fiber amplifiers (EDFAs) at 80~km span lengths. The
64~GBaud, 75~GHz grid is {used} as a {conservative} {single-carrier}
{reference} point. {
The 800G and 1.6T ZR/ZR+ interfaces run at substantially higher
single-carrier symbol rates ($\sim$118--130~GBaud for 800G and
$>$200~GBaud for 1.6T-class transponders). Because the transceiver
back-to-back SNR ceiling falls monotonically with baud
(Section~\ref{subsec:trx_ceiling}), 64~GBaud is an optimistic
operating point for the transceiver limit; the
full} symbol-rate dependence is swept explicitly in
Section~\ref{subsec:res_baudrate_heatmap}. Two
band placements are studied: a C-band grid starting at 1520~nm
(spanning 6~THz to $\sim$1567~nm), and an L-band grid starting
at 1572~nm (to $\sim$1623~nm) that overlaps the strong CO$_2$
absorption bands (Section~\ref{subsec:gas}). The 6-THz grids presume
extended (super-C/L-band) amplification; the residual per-amplifier gain
ripple that such wideband operation implies is covered by the
line-system penalty of Section~\ref{subsec:linesys}. Soft-decision forward
error correction (SD-FEC) is assumed at a pre-FEC bit-error rate (BER) threshold of
$2\times 10^{-2}$.

Table~\ref{tab:fiber_params} summarizes the fiber parameters. We
adopt 0.11~dB/km HCF attenuation as the design point throughout this
paper~\cite{Jasion2022,Petrovich2025}, 3.0~ps/(nm$\cdot$km) for the
chromatic dispersion, $\gamma=10^{-3}$\,W$^{-1}$\,km$^{-1}$ for
the Kerr nonlinear coefficient~\cite{Fokoua2023}, and a baseline
distributed IMI coefficient of $-55$~dB/km (the state-of-the-art is $-73$~dB/km~\cite{li_ofc2026}).

\begin{table}[!t]
	\caption{Fiber Parameters Used in This Analysis}
	\label{tab:fiber_params}
	\centering
	\small
	\begin{tabular}{lcc}
		\toprule
		\textbf{Parameter} & \textbf{SMF (G.652)} & \textbf{HCF (NANF)} \\
		\midrule
		Attenuation $\alpha$ (dB/km)                        & 0.20  & 0.11  \\
		Dispersion $D$ (ps/nm/km)                           & 17.0  & 3.0   \\
		Nonlinear coeff.\ $\gamma$ (W$^{-1}$km$^{-1}$)     & 1.4   & 0.001 \\
		PMD coeff.\ (ps/km$^{1/2}$)                         & 0.10  & 0.05  \\
		Splice/connector loss (dB)                          & 0.30  & 0.50  \\
		Mode-field diameter ($\mu$m)                        & 10.4  & 25.0  \\
		IMI coeff.\ $\kappa$ (dB/km)                        & ---   & $-55$ \\
		CO$_2$ line excess loss (dB/km)\textsuperscript{a}  & ---   & $0$   \\
		\bottomrule
		\multicolumn{3}{l}{\footnotesize\textsuperscript{a}Baseline assumes line avoidance / sealed core;}\\
		\multicolumn{3}{l}{\footnotesize worst-case channel-on-line studied in Sec.~\ref{subsec:res_gas} (up to 0.5~dB/km).}\\
	\end{tabular}
\end{table}
The transceiver and amplification configuration assumed throughout
is as follows: transmitter and local-oscillator laser linewidths
$\Delta\nu_{\text{TX}} = \Delta\nu_{\text{LO}} = 50$~kHz
(external-cavity laser, ECL); root-mean-square (RMS) timing jitter 100~fs;
total link PDL 2.0~dB, matching the Optical Internetworking Forum (OIF) 400ZR/800ZR worst-case
Implementation Agreement specification~\cite{OIF400ZR,OIF800ZR};
EDFA noise figure $F_n=5$~dB; an order-dependent probabilistic
constellation shaping (PCS) gain $G_{\text{PCS}}(M)$ anchored at the
PS-64-QAM sensitivity gain of $0.8$~dB reported by
Fehenberger~\textit{et~al.}~\cite{Fehenberger2016} and rising toward
the asymptotic $1.53$~dB additive white Gaussian noise (AWGN) limit at higher
orders~\cite{ChoWinzer2019} (Section~\ref{subsec:snr_th}); and a 1.0~dB
engineering margin~\cite{Pointurier2017}.

\emph{Transceiver classes.} Because the transceiver back-to-back SNR
(Section~\ref{subsec:trx_ceiling}) is decisive for high-order QAM, we
parameterize two classes. A high-performance class represents
best-in-class coherent front-ends: a reference $\text{ENOB}=5.5$ at
32~GBaud, consistent with the converter performance reported for
coherent-transceiver DACs/ADCs~\cite{Laperle2014,Murmann2026}; an ENOB
roll-off of $0.6$~bit/octave, a deliberately milder slope than the
$\sim$1~bit/octave implied by a constant-figure-of-merit reading of the
Walden/Murmann converter surveys~\cite{Walden1999,Murmann2026},
reflecting generational converter improvement; and $38$~dB
analog-front-end and $42$~dB Tx/Rx-nonlinearity back-to-back floors,
calibration parameters set so that the composite ceiling reproduces the
$\sim$10~dB bandwidth-induced penalty near 100~GBaud of the
frequency-dependent-ENOB analysis of Varughese
\textit{et~al.}~\cite{Varughese2018} and remains consistent with the
back-to-back SNRs realized in published high-order-QAM experiments
(Section~\ref{subsec:trx_ceiling}, Fig.~\ref{fig:trx_ceiling}). A
deployable pluggable class is $\sim$2--3~dB lower. Unless stated
otherwise the headline results use the high-performance class, so the
transceiver-limited conclusions below hold even under an optimistic
transceiver assumption. We also report results for $N_{\text{sc}}$
digital subcarriers, i.e., digital subcarrier multiplexing (DSCM), {an}
architecture {adopted in several} deployed {long-haul and metro coherent
transponders (the} 400ZR/800ZR {pluggables themselves use single-carrier}
{modulation).}

\emph{Gas-line assumption.} Air-filled HCF carries intrinsic CO$_2$
absorption lines; the strong lines lie in the L-band with weaker lines
in the C-band, where long-span HCF experiments have observed up to
$\sim$0.5~dB/km of excess loss at line
center~\cite{Wang2025_OL_CO2,saber2026_gla}. The
headline results assume a line-avoiding channel plan or a
sealed/evacuated core (no gas penalty); the worst-case channel-on-line
is treated separately in Section~\ref{subsec:res_gas}.

\emph{Launch power.} For SMF we cap the total WDM launch at $+26$~dBm
(high-power commercial booster); for HCF we adopt a high-power C-band
booster cap of $+40$~dBm total. This is supported both by the high aggregate launches
already used in nonlinearity-free HCF
demonstrations~\cite{Hong2024_HCF_terabit,Hong2025_HCF_JLT} and by the
energy-per-bit launch-power analysis of Sohanpal
\textit{et~al.}~\cite{Sohanpal2026_launch}, which shows that HCF's low
Kerr NLI pushes the optimum launch well above conventional SMF booster
levels. The operating launch is the Gaussian-noise (GN) model SNR optimum, clipped to this cap; as shown in
Section~\ref{subsec:res_pigtail}, the residual HCF NLI keeps this optimum
below the cap across the studied range, so the cap is not the binding
constraint.

For HCF we adopt a single Long-haul deployment scenario with
5~km HCF segment length, baseline HCF--HCF splice LP$_{11}$
excitation of $-30$~dB inferred from the recent fusion-splice
evaluation of Zhang \textit{et~al.}~\cite{Zhang2026}, an HCF-to-SMF
junction LP$_{11}$ excitation of $-35$~dB measured by
Suslov \textit{et~al.}\ with a graded-index mode-field
adapter~\cite{Suslov2021}, and an LP$_{11}$ DMA of 10~dB/km, consistent with reported HOMER values on the order of tens of dB over multi-km NANF/DNANF lengths \cite{Fokoua2023}. This
scenario is used for all numerical results
except for the splice-MPI sweep of Fig.~\ref{fig:imi_impact}, which
is evaluated at a tighter 2~km segment length to expose more
splice/segment combinations to the all-pairs MPI sum (the worst case
for splice-MPI accumulation; at 5~km segments splice MPI is even
smaller). The 2~km / 5~km segment choices reflect practical
terrestrial cable-deployment constraints, including duct geometry,
truckload capacity, and the inter-splice spacing required for
field-accessible repair and re-termination.

\subsection{SNR Budget}
\label{subsec:snr_budget}

We use an effective-SNR budget: a quantity is summed in the
SNR denominator only if it is a genuine additive noise power; effects
that are not literal noise powers (gas absorption, line-system
narrowing, PDL, the phase-recovery penalties, and the gas-notch inter-symbol interference (ISI))
are represented as SNR-equivalent penalties. The budget has
three cleanly separated groups, combined reciprocally. (i) The
link SNR collects the true optical noise powers,
\begin{equation}
	\text{SNR}_{\text{link}} = \frac{P_{\text{ch}}}{P_{\text{ASE}} + P_{\text{NLI}} + P_{\text{IMI,total}} + P_{\text{NLI,pig}}},
	\label{eq:snr_link}
\end{equation}
namely amplified spontaneous emission (ASE), fiber Kerr NLI, total IMI
(distributed plus discrete), and SMF-pigtail (and mode-field-adapter)
NLI. CO$_2$ gas-line absorption is not a noise power: a
channel overlapping a line carries extra distributed loss, which raises
the EDFA gain and hence the ASE on that channel (it enters $P_{\text{ASE}}$,
Section~\ref{subsec:gas}), not the denominator directly. (ii)
The transceiver back-to-back ceiling $\text{SNR}_{\text{TRx}}(M,R_s)$,
defined in Section~\ref{subsec:trx_ceiling}, is the pure B2B
quantity---converter ENOB at the operating sample rate, analog
bandwidth/anti-aliasing, and Tx/Rx nonlinearity only. It is
distance-independent. (iii) The remaining effects---laser phase
noise, EEPN (a link-dispersion$\times$LO-linewidth interaction, not a
B2B term), timing jitter, PDL, the network line-system penalty, and the
in-band gas-notch ISI---are SNR-equivalent penalties $\delta_i$, each an
equivalent noise floor referenced to $\text{SNR}_{\text{th}}(M)$. EEPN
and the line-system penalty are excluded from the B2B ceiling by
construction: both depend on link quantities (accumulated dispersion
and the amplifier/filter chain), so folding them into a
transceiver-only ceiling would misattribute link-dependent impairments
to the transceiver. The total SNR is
\begin{equation}
	\frac{1}{\text{SNR}_{\text{total}}} = \frac{1}{\text{SNR}_{\text{link}}} + \frac{1}{\text{SNR}_{\text{TRx}}(M,R_s)} + \sum_{i}\frac{1}{\text{SNR}_{i}},
	\label{eq:snr_total}
\end{equation}
where the reciprocal construction follows Klaus and
Winzer~\cite{Klaus2022_HCF}, here resolved into its components. QAM order
$M$ at distance $L$ and symbol rate $R_s$ is feasible when
\begin{equation}
	\text{SNR}_{\text{total}}\,[\mathrm{dB}] \;\ge\; \text{SNR}_{\text{th}}(M) - G_{\text{PCS}}(M) + \delta_{\text{margin}} + \delta_{\text{ISI}}^{\text{gas}},
	\label{eq:feasibility}
\end{equation}
where $\text{SNR}_{\text{th}}(M)$ is the BER-threshold $E_s/N_0$ per
polarization (Section~\ref{subsec:snr_th}), $G_{\text{PCS}}(M)$ is the
order-dependent shaping credit, $\delta_{\text{margin}}$ is the
engineering margin, and $\delta_{\text{ISI}}^{\text{gas}}$ is the
gas-notch ISI penalty (Section~\ref{subsec:gas}), added to the required
SNR in the way deterministic, partially equalizable waveform
distortions are conventionally carried in system budgets---as a bounded
dB penalty rather than as an accumulating noise power. The essential
difference from a dB-sum budget is
that Eq.~\eqref{eq:snr_total} caps the achievable SNR at
$\text{SNR}_{\text{TRx}}$: if the transceiver ceiling lies below the
requirement, the order is infeasible no matter how clean the link is.

\section{Impairment Models}
\label{sec:impairment_models}

\subsection{Theoretical SNR Requirement}
\label{subsec:snr_th}

For square $M$-QAM with Gray coding, the standard pre-FEC bit-error
rate is~\cite{Proakis2008,Yoon2000BEP_MQAM}
\begin{equation}
	\mathrm{BER} \approx \frac{4}{\log_2 M}\!\left(1-\frac{1}{\sqrt{M}}\right)\!Q\!\left(\sqrt{\frac{3\,(E_s/N_0)}{M-1}}\right)\!,
	\label{eq:ber_mqam_Q}
\end{equation}
where $Q(\cdot)$ is the standard normal Q-function and $E_s/N_0$ is
the per-symbol per-polarization SNR (equivalent to the optical
coherent $Q^2$ per polarization). Using $Q(x)=\tfrac{1}{2}\,
\mathrm{erfc}(x/\sqrt{2})$, an equivalent closed form in terms of
$\mathrm{erfc}(\cdot)$ is
\begin{equation}
	\mathrm{BER}\approx\frac{2}{\log_2 M}\!\left(1-\frac{1}{\sqrt{M}}\right)\!\mathrm{erfc}\!\left(\sqrt{\frac{3\,(E_s/N_0)}{2(M-1)}}\right)\!.
	\label{eq:ber_mqam_erfc}
\end{equation}
Numerically inverting Eq.~\eqref{eq:ber_mqam_Q} at
$\mathrm{BER}=2\times 10^{-2}$ (the SD-FEC pre-FEC threshold) gives
$\text{SNR}_{\text{th}}(E_s/N_0)$ per polarization of 6.25~dB (quadrature phase-shift keying, QPSK),
12.71~dB (16-QAM), 18.43~dB (64-QAM), 24.01~dB (256-QAM),
29.57~dB (1024-QAM), 32.35~dB (2048-QAM), and 35.14~dB (4096-QAM);
each step in QAM order adds about $5$--$6$~dB. These thresholds are
the per-polarization $E_s/N_0$ obtained directly from
Eq.~\eqref{eq:ber_mqam_Q} and are used as the
$\text{SNR}_{\text{th}}(M)$ entries in the feasibility test of
Eq.~\eqref{eq:feasibility}.

The probabilistic-shaping credit is order-dependent rather than a
fixed constant. We anchor $G_{\text{PCS}}(64)=0.8$~dB to the measured
PS-64-QAM sensitivity gain of Fehenberger~\textit{et~al.}~\cite{Fehenberger2016},
floor it near $0.4$~dB at 16-QAM, and let it rise toward the asymptotic
AWGN shaping-gain limit of $1.53$~dB at $\ge 1024$-QAM, consistent with
the order/SNR dependence reviewed by Cho and
Winzer~\cite{ChoWinzer2019}; numerically
$G_{\text{PCS}}(256)=1.20$~dB and $G_{\text{PCS}}(\ge 1024)=1.53$~dB.

\subsection{Transceiver Back-to-Back SNR Ceiling}
\label{subsec:trx_ceiling}

A real coherent transceiver delivers only a finite SNR even into an
ideal, infinite-OSNR link, and that ceiling falls as the symbol rate
rises. We emphasize that this ceiling is a parameterized model
anchored to reported converter and front-end trends---it is not
a measured back-to-back curve of a specific transceiver; its absolute
calibration shifts the exact crossover baud but not the qualitative
conclusion (a sensitivity to the calibration, and the two bracketing
transceiver classes, are given in
Section~\ref{subsec:res_trx_ceiling}). Representing residual
transceiver impairments as a lumped SNR floor that combines
reciprocally with the link SNR is established practice in
transceiver-noise-limited system analysis~\cite{Galdino2017_trxnoise,Klaus2022_HCF};
here that floor is resolved into rate-dependent components, modeled as
the parallel combination of three parameterized front-end floors. The
quantization floor uses the effective number of bits at the operating
symbol rate,
\begin{equation}
	\text{ENOB}(R_s) = \text{ENOB}_0 - s_b \log_2\!\left(R_s/R_0\right),
	\label{eq:enob_rate}
\end{equation}
\begin{equation}
	\text{SNR}_{\text{q}} = 6.02\,\text{ENOB}(R_s) + 1.76 - \Pi(M),
	\label{eq:snr_q}
\end{equation}
where $\text{ENOB}_0$ is quoted at $R_0=32$~GBaud, $s_b\approx0.6$~bit
per octave follows the Walden/Murmann converter
trend~\cite{Walden1999,Murmann2026}, and $\Pi(M)$ is the
constellation peak-to-average-power back-off
(Section~\ref{subsec:other_xcvr}). The analog-bandwidth and
anti-aliasing floor grows with baud,
$\text{SNR}_{\text{afe}}(R_s)=S_{\text{afe}}-10\log_{10}[1+(R_s/R_{\text{bw}})^2]$,
calibrated so the bandwidth penalty approaches $\sim$10~dB near
100~GBaud, consistent with reported high-baud transceiver
behavior~\cite{Varughese2018}; the Tx/Rx nonlinearity floor scales
with PAPR, $\text{SNR}_{\text{nl}}=S_{\text{nl}}-\Pi(M)$. These three
pure back-to-back floors combine in parallel,
\begin{equation}
	\frac{1}{\text{SNR}_{\text{TRx}}} = \frac{1}{\text{SNR}_{\text{q}}} + \frac{1}{\text{SNR}_{\text{afe}}} + \frac{1}{\text{SNR}_{\text{nl}}}.
	\label{eq:trx_ceiling}
\end{equation}
The distance/laser-dependent phase-noise, EEPN, jitter, PDL, and
line-system floors of
Sections~\ref{subsec:phase_noise}--\ref{subsec:other_xcvr} are
not part of this B2B ceiling; they enter the budget as the
SNR-equivalent penalties $\{\text{SNR}_i\}$ of
Eq.~\eqref{eq:snr_total}. For the high-performance class
($\text{ENOB}_0=5.5$, $S_{\text{afe}}=38$~dB, $S_{\text{nl}}=42$~dB) the
resulting ceiling (evaluated at the 256-QAM back-off) is $\approx$28.8~dB
at 32~GBaud, falling to $\approx$25.3~dB at 64~GBaud, $\approx$23.0~dB at
96~GBaud, and $\approx$21.2~dB at 128~GBaud; the deployable class is
$\sim$2.5--3~dB lower (Table~\ref{tab:ceiling}). Because the 1024- and
2048-QAM thresholds are $29.6$ and $32.4$~dB, the
ceiling alone forbids these orders---and, a fortiori, all higher
ones---at 64~GBaud and above, independent of
the fiber; even 256-QAM (24.0~dB) requires the high-performance class
above $\sim$45~GBaud. Extending the same model to low baud provides an
external check on its calibration: at 20~GBaud the high-performance
ceiling evaluated at the 1024-QAM back-off is $\approx$30.3~dB, above
the $\approx$29.0~dB net PS-1024-QAM requirement (threshold minus
shaping credit plus margin), and the full budget including the 5-km
link terms indeed returns 1024-QAM as the maximum feasible order at
20~GBaud over 5~km---matching the DNANF demonstration of
Fan~\textit{et~al.}~\cite{Fan2026_1024QAM_DNANF}. At 16~GBaud the
ceiling rises to $\approx$31.1~dB, while the 2048-QAM
requirement is met only below $\sim$10~GBaud
(Fig.~\ref{fig:trx_ceiling}), consistent with the low symbol rates and
research-grade instrumentation of the published uniform-constellation
SMF ultra-high-QAM records~\cite{Beppu2015_2048QAM,Terayama2018}
(deeply shaped PS variants, whose effective required SNR lies well
below the nominal-order threshold, reach somewhat higher
baud~\cite{Chen2019_4096QAM30G}).

\begin{table}[!t]
	\caption{Parameterized Transceiver B2B SNR Ceiling vs.\ Symbol Rate (256-QAM back-off)}
	\label{tab:ceiling}
	\centering
	\small
	\begin{tabular}{lccccc}
		\toprule
		Symbol rate (GBaud)        & 16 & 32 & 64 & 96 & 128 \\
		\midrule
		High-performance (dB)      & 31.4 & 28.8 & 25.3 & 23.0 & 21.2 \\
		Deployable (dB)            & 28.4 & 26.0 & 22.6 & 20.2 & 18.4 \\
		\bottomrule
	\end{tabular}
\end{table}

\subsection{Laser Phase Noise and Carrier-Phase Estimation}
\label{subsec:phase_noise}

The transmitter and local-oscillator (LO) lasers each contribute an
independent Wiener phase process. Because the convolution of two
independent Wiener processes with linewidths $\Delta\nu_{\text{TX}}$
and $\Delta\nu_{\text{LO}}$ is itself a Wiener process with linewidth
equal to the sum of the two (see, e.g., the carrier-recovery
derivations of Pfau~\textit{et~al.}~\cite{Pfau2009} and
Savory~\cite{Savory2008}), the
equivalent linewidth seen by the carrier-phase estimator is
$\Delta\nu_{\text{TX}}+\Delta\nu_{\text{LO}}$. We nonetheless keep
the two linewidths as separate parameters in the analysis so that any
mismatch between transmitter and LO can be made explicit. The CPE
block partially corrects for the slowly-varying part of the combined
process; the residual phase-noise penalty depends on the CPE
algorithm and on the constellation density.

We model three CPE classes parameterized by the dimensionless
linewidth-symbol-period product
$\Delta\nu\!\cdot\!T_s$~\cite{Pfau2009,Savory2008,Mazur2019}, where
$\Delta\nu=\Delta\nu_{\text{TX}}+\Delta\nu_{\text{LO}}$ and $T_s$ is
the symbol period. The 1~dB-penalty tolerance for QAM order $M$ and
a given CPE class is denoted $\tau_{1\text{dB}}(M)$. The penalty
model used throughout is the soft quadratic approximation
\begin{equation}
	\delta_{\text{PN}} = \left(\frac{\Delta\nu\,T_s}{\tau_{1\text{dB}}(M)}\right)^{\!2}\,\mathrm{dB},
	\label{eq:pn_model}
\end{equation}
in which the penalty is normalized to $\tau_{1\text{dB}}(M)$ so that
$\delta_{\text{PN}}=1$~dB at $\Delta\nu\!\cdot\!T_s=\tau_{1\text{dB}}(M)$
by construction. Equation~\eqref{eq:pn_model} reproduces the
quadratic small-signal behavior of the Pfau BPS variance
analysis~\cite{Pfau2009} (where the dominant residual phase-noise
variance scales as $\Delta\nu\,T_s$ multiplied by the CPE block
length) and provides a smooth interpolation that matches reported
penalty curves for pilot-aided
CPE~\cite{Sasai2020,Mazur2019,Fang2024}.

\subsubsection{Blind phase search (BPS), $\tau^{\mathrm{BPS}}_{1\text{dB}}(M)$}
We use the canonical tolerances from Pfau~\textit{et~al.} for $N=32$
block size and a BER target of $10^{-3}$~\cite{Pfau2009}:
$4.1\times10^{-4}$ (QPSK), $1.4\times10^{-4}$ (16-QAM),
$4.0\times10^{-5}$ (64-QAM), and $8.0\times10^{-6}$ (256-QAM). The
value used here for 1024-QAM ($1.5\times10^{-6}$) extends the Pfau
scaling and is not directly experimentally validated; it should be
read as a model value, not a measured one.

\subsubsection{Modern pilot-aided CPE, $\tau^{\mathrm{mod}}_{1\text{dB}}(M)$}
Pilot-aided CPE inserts known reference symbols and removes the
constellation-density-dependent decision feedback that limits
BPS~\cite{Sasai2020,Mazur2019,Shi2024_pilot}. We model the relative
tolerance improvement as
\begin{equation}
	G_{\text{mod}}(M) \equiv \frac{\tau^{\mathrm{mod}}_{1\text{dB}}(M)}{\tau^{\mathrm{BPS}}_{1\text{dB}}(M)} = 1 + 0.88\cdot \frac{\log_2 M}{8},
	\label{eq:G_modern}
\end{equation}
with $G_{\text{mod}}(256)=1.88$. The 256-QAM anchor
$\tau^{\mathrm{mod}}_{1\text{dB}}(256) \approx 1.5\times 10^{-5}$
implied by Eq.~\eqref{eq:G_modern} is calibrated to roughly match
the pilot-aided-CPE tolerance trends that Sasai
\textit{et~al.}~\cite{Sasai2020} report for PDM PS-1024-QAM and
US-256-QAM at 16~GBaud, extrapolated to a representative 32~GBaud,
PS-256-QAM operating point. We emphasize that
Eq.~\eqref{eq:G_modern} is a model assumption; the published
pilot-CPE literature reports tolerance values for individual
modulation orders and baud rates but does not give a closed-form
scaling with~$M$. The expression should be read as a smooth
interpolation that produces the right qualitative trend between
the published anchors at 64-QAM and 256-QAM.

\subsubsection{Wireless-style residual-carrier-modulation (RCM) CPE, $\tau^{\mathrm{RCM}}_{1\text{dB}}(M)$}
HCF's near-linear channel preserves pilot-tone signal-to-noise ratio
better than SMF, which makes residual-carrier-modulation
schemes---deliberate IQ modulator bias offset to leave a residual
carrier that beats with the signal in the digital
domain~\cite{Fang2024}---particularly attractive. Following the same
calibration philosophy we model
\begin{equation}
	G_{\text{RCM}}(M) \equiv \frac{\tau^{\mathrm{RCM}}_{1\text{dB}}(M)}{\tau^{\mathrm{BPS}}_{1\text{dB}}(M)} = 1 + 7.6\cdot \frac{\log_2 M}{8},
	\label{eq:G_wireless}
\end{equation}
with $G_{\text{RCM}}(256)=8.6$, matching the 256-QAM operating point
$\Delta\nu\!\cdot\!T_s=6.89\times10^{-5}$ reported
in~\cite{Fang2024}. The RCM technique has been
demonstrated up to PS-256-QAM only, and its extension to 1024-QAM in
our analysis is purely an interpolation. 

\subsection{ENOB, EEPN, Timing Jitter, and PDL}
\label{subsec:other_xcvr}

\textit{ENOB.} The classical signal-to-quantization-noise ratio
(SQNR) ceiling for an ideal $b$-bit converter driven by a full-scale
sinusoid is $\text{SQNR}=6.02\cdot\text{ENOB}+1.76$~dB, derived from
Bennett's quantization-noise model and widely used in ADC-survey
work~\cite{Walden1999}. For an $M$-QAM signal the effective SQNR is
reduced by the back-off required by the constellation peak-to-average
power ratio (PAPR). Adopting the convention in which $d$ is half the
minimum distance between adjacent symbols, so that the $M$-QAM
symbols lie at $(\pm 1, \pm 3,\ldots, \pm(\sqrt{M}-1))\,d$ on each
quadrature axis, the corner symbols sit at $\sqrt{2}(\sqrt{M}-1)\,d/\sqrt{(2/3)(M-1)\,d^2}$
times the RMS amplitude, giving an analytical pure-constellation
PAPR of $3(\sqrt{M}-1)/(\sqrt{M}+1)$ in the linear domain
($\approx 2.55$, $3.68$, $4.23$, $4.50$, and $4.64$~dB for $M = 16$,
$64$, $256$, $1024$, and $4096$, respectively). The PAPR values
used in the SQNR formula are referenced to the $3$~dB peak-to-RMS
of the sinusoidal baseline already implicit in
$6.02\cdot\text{ENOB}+1.76$, i.e., they correspond to the back-off
relative to a full-scale sinusoid; we take $\approx 0.5$~dB for
QPSK (near-constant modulus, dominated by spectral-shaping
overshoot) and $3.0+0.25\cdot\max(0,\log_2 M-4)$~dB for $M\ge 16$,
which combines the analytical constellation PAPR above with the
$\sim$1~dB pulse-shaping contribution typical of root-raised-cosine
spectral shaping ($\alpha\approx 0.1$--$0.2$) and matches the
$\sim$4.5--5~dB band reported for $M\ge 1024$ in pulse-shaped
optical-coherent links. The penalty form
\begin{equation}
	\delta_{\text{ENOB}} = 10\log_{10}\!\left(1+\frac{\text{SNR}_{\text{th,lin}}(M)}{\text{SQNR}_{\text{eff,lin}}}\right)
\end{equation}
is the standard Gaussian noise-power-addition identity, in which
the quantization noise adds incoherently to the signal-domain
noise floor. The frequency-dependent ENOB framework for M-QAM
optical links of Varughese~\textit{et~al.}~\cite{Varughese2018}
provides simulation-validated OSNR penalty curves that this closed
form approximates. The reference SNR in the numerator is the
BER-threshold $E_s/N_0$ of the chosen QAM order; this keeps the ENOB
penalty an explicit function of the modulation format alone. The
quantization floor enters the transceiver ceiling of
Eq.~\eqref{eq:trx_ceiling} as $\text{SNR}_{\text{q}}$, where it is
combined in parallel with the other front-end floors rather than added
as a dB penalty: quantization noise is, to first order, an additive
noise power independent of the other floors~\cite{Walden1999}, so
power-domain (reciprocal-SNR) combination is the physically consistent
composition and is the treatment used in transceiver-noise-limited
system analysis~\cite{Galdino2017_trxnoise,Klaus2022_HCF}.

\textit{EEPN.} The LO laser's phase noise interacts with the digital
chromatic-dispersion equalizer's delay spread to add a per-symbol
noise term whose noise-to-signal variance is, after Shieh and
Ho~\cite{Shieh2008,Xie2009},
\begin{equation}
	\sigma^2_{\text{EEPN}} = \frac{\pi\,\lambda^2\,|D|\,L\,\Delta\nu_{\text{LO}}\,R_s}{c}\,,
	\label{eq:eepn_var}
\end{equation}
where $\Delta\nu_{\text{LO}}$ is the LO linewidth (only; the
transmitter linewidth does not enter EEPN), $D\,L$ is the accumulated
dispersion (with $D$ and $L$ in matching SI units), $R_s = 1/T_s$
is the symbol rate, $\lambda$ is the wavelength, and $c$ is the
speed of light. With {DSCM---an} architecture {adopted in several} deployed
coherent digital signal processing (DSP) application-specific integrated circuits (ASICs)~\cite{Sun2020_DSCM}, {though not in the
single-carrier 400ZR/800ZR pluggables---each} of
$N_{\text{sc}}$ subcarriers carries $R_s/N_{\text{sc}}$ and sees a
correspondingly smaller chromatic-dispersion (CD) equalizer group-delay spread, so to first
order $\sigma^2_{\text{EEPN}}$ scales as $R_s/N_{\text{sc}}$; we
therefore replace $R_s\!\to\!R_s/N_{\text{sc}}$ in
Eq.~\eqref{eq:eepn_var} ($N_{\text{sc}}=1$ recovers the single-carrier
result). The associated noise-floor SNR is
$\text{SNR}_{\text{EEPN}}\equiv 1/\sigma^2_{\text{EEPN}}$ and enters the
transceiver ceiling of Eq.~\eqref{eq:trx_ceiling}.

\textit{Timing jitter} contributes a noise floor set by the
signal mean-square bandwidth, not the full symbol rate:
$\text{SNR}_{\text{jit}}^{-1}=(2\pi\sigma_t)^2\langle f^2\rangle(1-\rho)$
with $\langle f^2\rangle\!\approx\!(R_s^2/12)(1+\beta^2)$ for a
near-Nyquist root-raised-cosine signal of roll-off $\beta$ and $\rho$
the Tx/Rx common-clock jitter correlation~\cite{Proakis2008}. For a
single full-scale tone at $R_s$ this expression reduces to the
classical aperture-jitter limit $(2\pi R_s\sigma_t)^2$~\cite{Walden1999};
weighting by the actual near-Nyquist signal spectrum lowers the floor
by $12/(1+\beta^2)$ ($\approx$10.8~dB at $\beta=0.1$). At the 64~GBaud,
100~fs operating point the jitter penalty is a few tenths of a dB
for $M\le 256$, and the frequency-dependent aperture-jitter component
at very high baud is noted in Section~\ref{subsec:limitations}.
\textit{PDL} contributes a polarization-tributary imbalance penalty
that grows mildly with QAM order, since denser constellations have
tighter decision boundaries and are more sensitive to per-polarization
SNR asymmetry. At the 2~dB OIF-spec total link PDL we model this
contribution as staying below $\sim$1~dB for $M\le 256$ and rising
to a few dB for $M\ge 1024$; the precise scaling is a modeling
assumption used here rather than a verified literature value, and
the budgets reported in Section~\ref{sec:results} treat PDL as a
fixed engineering contribution at this representative level.

\subsection{Fiber Kerr Nonlinear Interference}
\label{subsec:nli}

Fiber Kerr NLI is computed with the enhanced Gaussian-noise (eGN)
model~\cite{Poggiolini2012,Poggiolini2015_eGN}. The
per-span Nyquist-WDM efficiency is
\begin{equation}
	\eta = \frac{8}{27}\,\frac{\gamma^2 L_{\text{eff}}^2}{\pi |\beta_2|}\,\mathrm{arcsinh}\!\left(\frac{\pi^2}{2}|\beta_2| L_{\text{eff}} B_{\text{WDM}}^2\right)\!,
	\label{eq:eta_nli}
\end{equation}
with $L_{\text{eff}}=(1-e^{-\alpha L_s})/\alpha$ and
$\beta_2 = -D\lambda^2/(2\pi c)$. The total NLI noise in the channel
under test is $P_{\text{NLI}} = \eta\,P_{\text{ch}}^3\,N_s^{1+\epsilon}$
with $\epsilon\approx 0.05$, and the modulation-format-dependent
eGN correction~\cite{Poggiolini2015_eGN} is applied to $\eta$ for
the QAM orders considered here.

The optimum per-channel launch power that maximizes
$P_{\text{ch}}/(P_{\text{ASE}} + \eta_{\text{tot}} P_{\text{ch}}^3)$
is
\begin{equation}
	P_{\text{opt}} = \left(\frac{P_{\text{ASE}}}{2\,\eta_{\text{tot}}}\right)^{\!1/3}\!,
	\label{eq:popt}
\end{equation}
where $\eta_{\text{tot}}$ aggregates all cubic-power NLI sources
(fiber and pigtail). The IMI term, which is linear in
$P_{\text{ch}}$, cancels in the derivative of the SNR and does not
move $P_{\text{opt}}$: writing the per-channel SNR as
$P_{\text{ch}}/(P_{\text{ASE}}+\eta_{\text{tot}}P_{\text{ch}}^3+a\,P_{\text{ch}})$
with $a$ the IMI coefficient, setting
$\mathrm{d}\,\mathrm{SNR}/\mathrm{d}P_{\text{ch}}=0$ eliminates the
linear-in-$P_{\text{ch}}$ contribution and leaves the standard
GN-model optimum of Eq.~\eqref{eq:popt}. The IMI term does,
however, reduce the achievable SNR at the optimum.

\subsection{Inter-Modal Interference}
\label{subsec:imi}

We decompose IMI into three components, all summed incoherently in
the noise-power domain.

\subsubsection{Distributed fiber IMI}
Following Poggiolini and Poletti~\cite{Poggiolini2022_HCF},
\begin{equation}
	P_{\text{IMI,fiber}} = \kappa\,P_{\text{ch}}\,L_{\text{total}},
	\label{eq:imi_fiber}
\end{equation}
with $\kappa$ the linear distributed IMI coefficient (per km).

\subsubsection{Discrete splice MPI: DMA-aware all-pairs sum}
At each splice or connector the mode-field discontinuity excites the
LP$_{11}$ mode with linear power coupling $\kappa\ll 1$. For MPI to be generated on the fundamental mode
the excited LP$_{11}$ must propagate to a downstream coupling site
and convert back to LP$_{01}$ there. Within the coupled-mode
framework of Mlejnek~\textit{et~al.}~\cite{Mlejnek2015} (later
validated experimentally by Downie~\textit{et~al.}~\cite{Downie2017}),
the single-path MPI power transferred between two coupling sites
separated by $L_{\text{seg}}$ with one-way LP$_{11}$ couplings
$\kappa_{\text{in}}$ at the entry site and $\kappa_{\text{out}}$ at
the exit site is the product of (i)~the LP$_{01}\to$LP$_{11}$ coupling
fraction at the first site, (ii)~the DMA-induced attenuation of
LP$_{11}$ over $L_{\text{seg}}$, and (iii)~the
LP$_{11}\to$LP$_{01}$ re-coupling at the second site, giving
\begin{equation}
	\mathrm{MPI}_{\text{path}} = \kappa_{\text{in}}\,\kappa_{\text{out}}\,10^{-\mathrm{DMA}\,L_{\text{seg}}/10},
	\label{eq:mpi_path}
\end{equation}
where DMA is the LP$_{11}$ differential modal attenuation relative
to LP$_{01}$, in dB/km. The same paper shows that a span
containing $N_s$ splices accumulates MPI from all unordered
pairs $(i,j)$ of coupling sites within the span, not just nearest
neighbors. With $\zeta\equiv 10^{-\mathrm{DMA}\,L_{\text{seg}}/10}$
denoting the DMA attenuation factor per segment,
the per-span sum over all such pairs whose entry and exit splice are
separated by $k$ segments is
\begin{equation}
	S(N_s,\zeta) = \sum_{k=1}^{N_s-1}(N_s-k)\,\zeta^k,
	\label{eq:mpi_sum_general}
\end{equation}
where the factor $(N_s-k)$ counts the number of distinct splice-pair
combinations separated by exactly $k$ segments. Equation~%
\eqref{eq:mpi_sum_general} admits the closed form
\begin{equation}
	S(N_s,\zeta) = \zeta\,\frac{N_s-1-N_s\zeta+\zeta^{N_s}}{(1-\zeta)^2}\quad(\zeta\neq 1),
	\label{eq:mpi_sum}
\end{equation}
obtained from the geometric series $\sum_{k=1}^{N_s-1}\zeta^k$ and
its first derivative. The full per-span splice MPI is then
$\kappa_{\text{splice}}^2\,S(N_s,\zeta)$, and analogous closed forms
apply to junction-to-splice and junction-to-junction paths. DMA
values reported in the literature span 1--20~dB/km for early NANFs
to $>$20~dB/km for DNANF~\cite{Poletti2014,Jasion2022},
with resonant designs reaching $>$100~dB/km.

\subsubsection{Longitudinal back-reflection MPI}
At each glass--air boundary Fresnel back-reflection produces a
backward LP$_{01}$ echo. With combined angled-interface and
anti-reflection coating the per-splice return loss measured by
Suslov~\textit{et~al.} is below $-60$~dB at
1550~nm~\cite{Suslov2022_backref,Shi2024_connector}, which renders
this contribution negligible at the dB level for our parameters; we
keep the term in the budget for completeness.

\subsection{SMF Pigtail NLI}
\label{subsec:pigtail}

Every EDFA in an HCF link has SMF pigtails on both sides connected to
HCF via mode-field adapters; in fact, every other line-system element
(boosters, ROADMs, OADMs, monitor taps) is also built around SMF and
adds further pigtail length, but the dominant pigtail NLI contribution
comes from the EDFA output pigtail because that is where the
signal power is highest and NLI scales as the cube of the launched
power.

The output pigtail of each amplifier carries the signal at the full
launch power $P_{\text{ch}}$. The input pigtail of the next
amplifier carries the same signal but attenuated by one HCF span loss,
i.e.\ at power $P_{\text{ch}}\cdot 10^{-\alpha_{\text{HCF}}L_s/10}$.
Because NLI scales as $P^3$ in the GN
model~\cite{Poggiolini2012}, the input-pigtail contribution is
suppressed by a factor $10^{-3\alpha_{\text{HCF}}L_s/10}\approx
2.3\times 10^{-3}$ for our $\alpha_{\text{HCF}}=0.11$~dB/km,
$L_s=80$~km baseline (an HCF span attenuates the signal by
$\sim$8.8~dB, which becomes $\sim$26~dB after the cubic). Adding
the two contributions over $N_s$ spans gives
\begin{equation}
	P_{\text{NLI,pig}} = N_s\!\left(1+10^{-3\alpha_{\text{HCF}} L_s/10}\right)\eta_{\text{SMF}}(L_{\text{pig}})\,P_{\text{ch}}^3,
	\label{eq:pigtail_nli}
\end{equation}
where $\eta_{\text{SMF}}(L_{\text{pig}})$ is the GN-model coefficient
of Eq.~\eqref{eq:eta_nli} evaluated for short SMF segments of
effective length $L_{\text{eff}}\approx L_{\text{pig}}$ (because
$L_{\text{pig}}\ll 1/\alpha_{\text{SMF}}\approx 21.7$~km).
Eq.~\eqref{eq:pigtail_nli} uses a linear span-count factor $N_s$,
in contrast with the $N_s^{1+\epsilon}$ used for the fiber NLI in
Section~\ref{subsec:nli}: the small coherence-build-up exponent
$\epsilon\approx 0.05$ reflects partial inter-span phase coherence
that accumulates only over long, dispersion-spread fiber spans, and
vanishes ($\epsilon_{\text{pig}}\approx 0$) for the
meter--hundred-meter pigtails considered here because they
accumulate negligible chromatic dispersion. Substituting
Eq.~\eqref{eq:pigtail_nli}
into Eq.~\eqref{eq:popt} with the per-link aggregate
\begin{equation}
	\eta_{\text{tot}}^{\text{(link)}} = \eta_{\text{fiber}}\,N_s^{1+\epsilon} + \eta_{\text{SMF}}(L_{\text{pig}})\,N_s\!\left(1+10^{-3\alpha_{\text{HCF}}L_s/10}\right),
	\label{eq:eta_tot_link}
\end{equation}
so that $P_{\text{NLI,total}} = \eta_{\text{tot}}^{\text{(link)}}\,P_{\text{ch}}^3$
gives the pigtail-aware optimum launch power. The
$N_s^{1+\epsilon}$ on the fiber term and $N_s$ on the pigtail term
make explicit that the two contributions accumulate with different
coherence factors.

\emph{Mode-field-adapter (MFA) taper NLI.} Tapered-SMF MFAs contain a
short ($\sim$cm) small-effective-area section of locally elevated
$\gamma$ that carries the full launch power; at HCF's high launch this
adds Kerr NLI. We append it to Eq.~\eqref{eq:pigtail_nli} as a
$\sim$cm high-$\gamma$ segment at each output-MFA site. It is sub-dB at
the operating launch but, scaling as $P^3$, becomes material at the
highest-power corner; graded-index-type adapters that avoid the taper
remove the term. The MFA coefficient is estimated, not measured.

\subsection{Gas-Line (CO$_2$) Absorption}
\label{subsec:gas}
Air-filled HCF carries intrinsic CO$_2$ absorption from its near-IR
combination bands. We model the actual line structure: a comb of pressure-broadened Lorentzian lines on the
$P$/$R$ rotational branches (spacing $\sim$23~GHz, FWHM $\sim$1.5~GHz,
Boltzmann strength envelope) of three bands---a weak band near
1540~nm in the C-band and strong bands near 1573 and 1606~nm in
the L-band (the 1572~nm line is the well-known CO$_2$ DIAL
line)~\cite{Wang2025_OL_CO2,Sillekens2026_GLA}. The comb is anchored in
absolute strength to the up-to-$0.5$~dB/km line-center excess loss
measured by Wang \textit{et~al.}~\cite{Wang2025_OL_CO2}. A finite
($R_s$-wide) channel overlaps only $\sim$1--2 of the narrow lines, so the
excess loss it experiences is the channel-averaged comb value
$g(\nu_{\text{ch}},R_s)$, well below the line-center peak and
baud-dependent (a wider channel averages over more of the comb).

The penalty has two physically distinct parts.
\emph{(a) Excess-loss$\rightarrow$ASE.} A channel-on-line carries extra
distributed loss $g$; the EDFA must supply that gain per span, raising
its ASE. This enters $P_{\text{ASE}}$ in Eq.~\eqref{eq:snr_link} (span
gain $10^{(\alpha+g)L_s/10}$), not the denominator as a separate
noise power. On the IMI-limited HCF link this OSNR hit is largely
masked (IMI sits well above the ASE floor), so excess loss alone is a
poor description of the measured impact.
\emph{(b) In-band notch ISI.} The dominant channel-on-line penalty
measured in HCF is the inter-symbol interference from the narrow spectral
notch, which a linear equalizer only partly inverts (it enhances noise
across the notch)~\cite{Sillekens2026_GLA}. Being a per-channel signal
distortion rather than an added noise floor, it is not masked by
IMI. We represent it as a bounded SNR-equivalent penalty added to the
required SNR,
\begin{equation}
	\delta_{\text{ISI}}^{\text{gas}} = \delta_{\max}\!\left(1 - e^{-gL/A_0}\right),
	\label{eq:gas_isi}
\end{equation}
using the accumulated in-band loss $gL$ as a notch-depth proxy. The
form captures the two limits of the physics: for a shallow notch the
residual ISI power grows in proportion to the notch depth, and hence
to $gL$; for a deep notch the adaptive equalizer bounds the residual
penalty (deeper attenuation of an already-opaque notch changes the
equalized response little), so $\delta$ saturates at $\delta_{\max}$.
{This saturating form is the finite-tap real-time equalizer directly: a
bounded-tap equalizer cannot fully invert a deep, narrow notch, so its
residual plateaus. A hard limit follows: once the accumulated notch
drives the required SNR above what the transceiver delivers, the channel
is undecodable---an effective notch-depth threshold beyond which
real-time decoding fails---so the budget returns no feasible order for
it.}
The two calibration constants, $\delta_{\max}\approx5.5$~dB and
$A_0\approx6$~dB, are anchored to the equalization and
pre-equalization results of Sillekens and
Sohanpal~\cite{Sillekens2026_GLA} (a few-dB notch leaves a $\sim$1--3~dB
residual after realistic equalization), so that with part~(a) the worst
L-band channel-on-line approaches the $\sim$6~dB penalty reported in
\cite{Wang2025_OL_CO2}; this calibration freedom is carried in
Table~\ref{tab:limitations}. Setting $g=0$ (line avoidance, or a
sealed/evacuated core, which removes the gas entirely) zeroes both parts;
spectral pre-emphasis or DSCM spectral avoidance can recover much of the
penalty for an unsealed core~\cite{Wang2025_OL_CO2,saber2026_gla}.

\subsection{Line-System Penalty}
\label{subsec:linesys}
Residual EDFA gain ripple/tilt (after gain-flattening) accumulating over
the amplifier chain, and ROADM/wavelength-selective switch (WSS) filter narrowing growing with the
number of cascaded filtering elements, both erode the effective SNR. We
include an optional line-system penalty applied identically to SMF and
HCF; the baseline is a point-to-point amplified line (no ROADMs), and a
worked multi-ROADM example is given in Section~\ref{subsec:limitations}.

\section{Numerical Results}
\label{sec:results}

All results below assume the system configuration of
Section~\ref{sec:system_model} unless noted. We first examine the
individual impairments in isolation (EEPN, phase noise, IMI, pigtail
NLI), then build up the integrated SNR budget at 100~km, and finally
present the system-level QAM-reach trade.

\begin{figure}[!t]
	\centering
	\includegraphics[width=0.9\columnwidth]{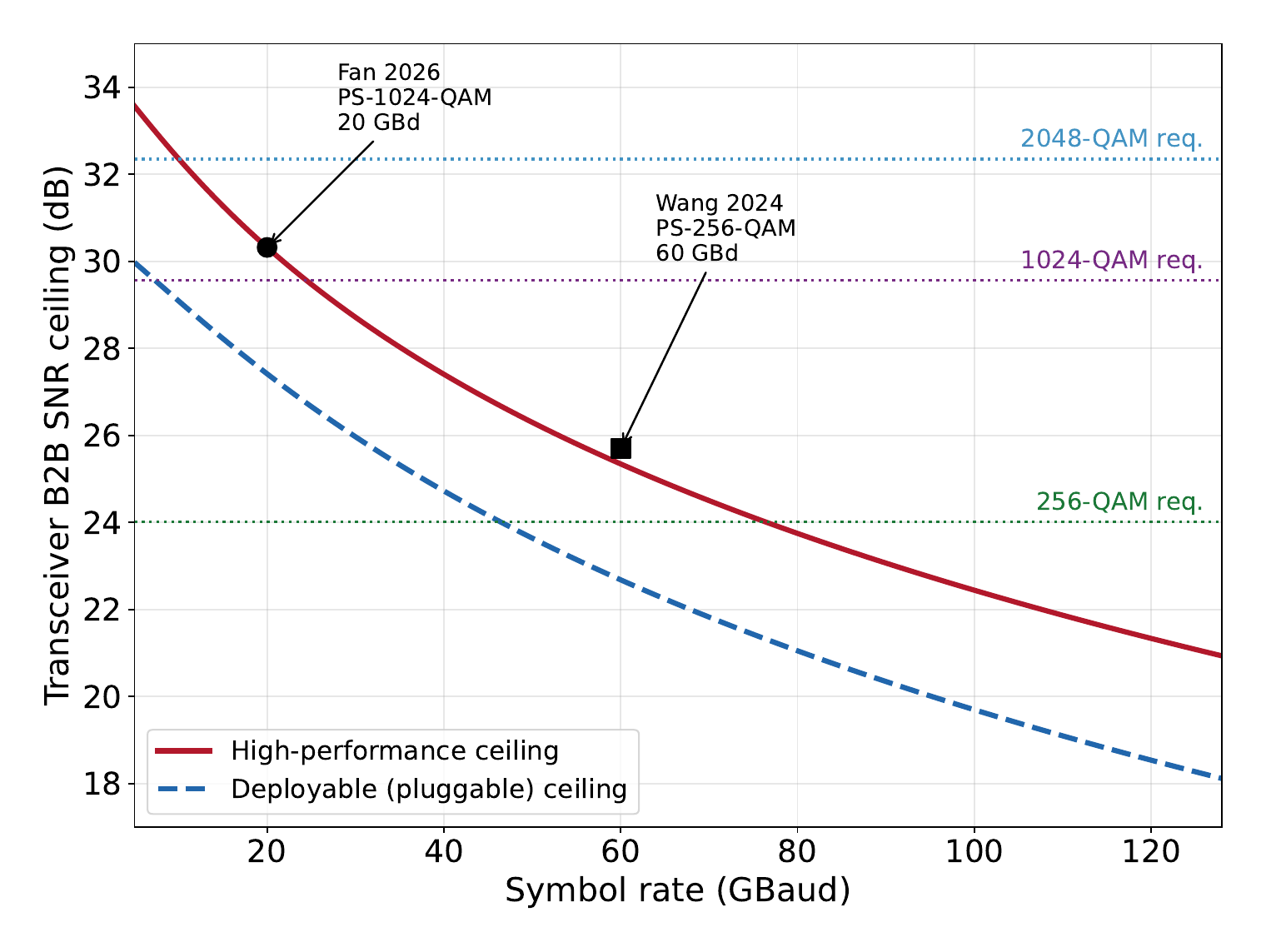}
	\vspace{-0.55cm}
	\caption{Transceiver back-to-back SNR ceiling
		$\text{SNR}_{\text{TRx}}(R_s)$ versus symbol rate for the
		high-performance and deployable classes, with the required-SNR
		lines for 256/1024/2048-QAM. The ceiling falls with baud and
		crosses the 1024-QAM requirement near 24~GBaud; the published HCF
		PS-1024-QAM (20~GBaud)~\cite{Fan2026_1024QAM_DNANF} and
		PS-256-QAM (60~GBaud)~\cite{Wang2024_NANF} demonstrations sit at
		the corresponding boundaries.}
	\label{fig:trx_ceiling}
	\vspace{-0.55cm}
\end{figure}

\begin{figure}[!t]
	\centering
	\includegraphics[width=0.9\columnwidth]{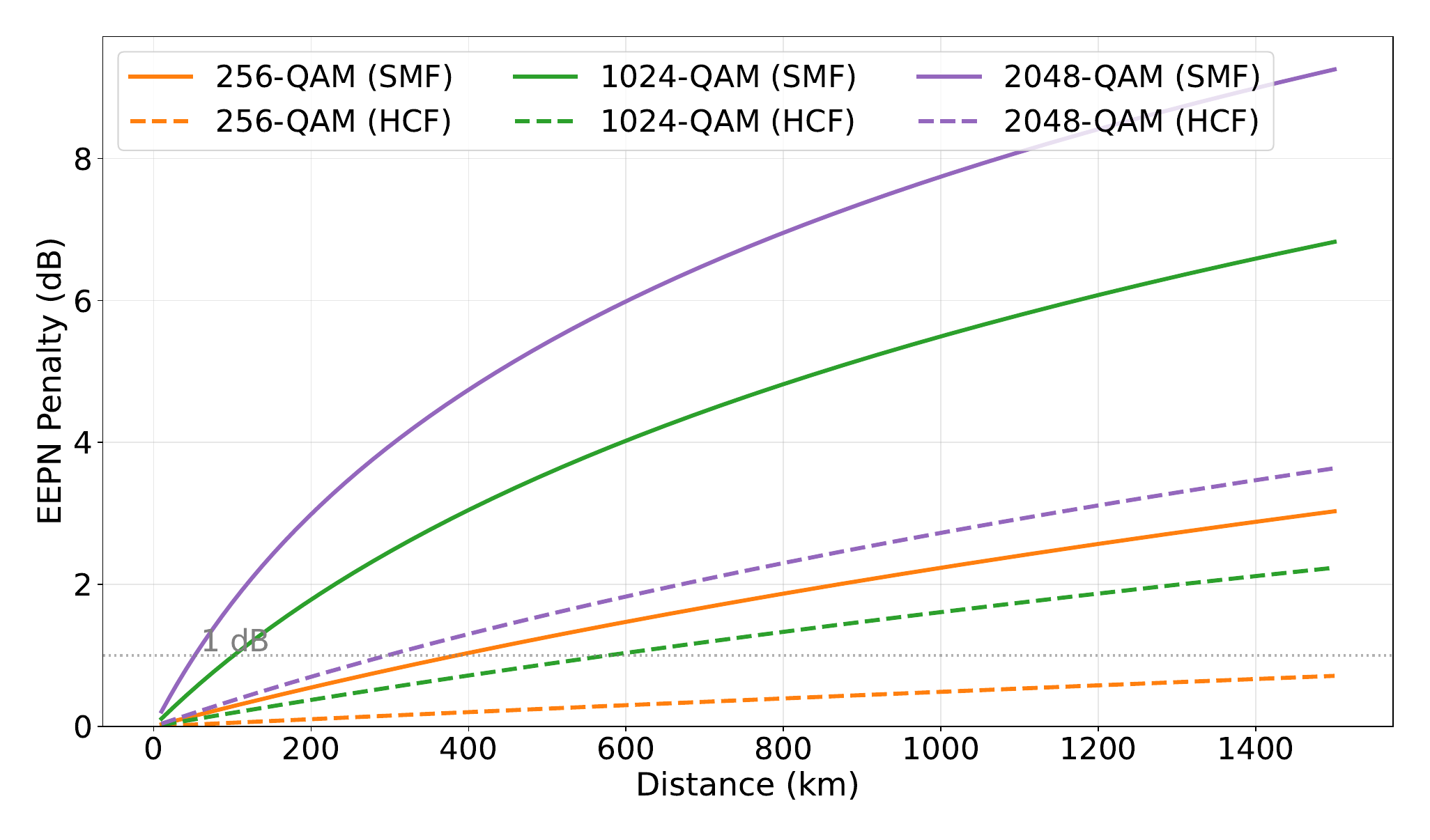}
	\vspace{-0.55cm}
	\caption{EEPN penalty versus distance for SMF and HCF across QAM
		orders.}
	\label{fig:eepn_comparison}
	\vspace{-0.55cm}
\end{figure}

\begin{figure*}[!t]
	\centering
	\includegraphics[width=0.63\textwidth]{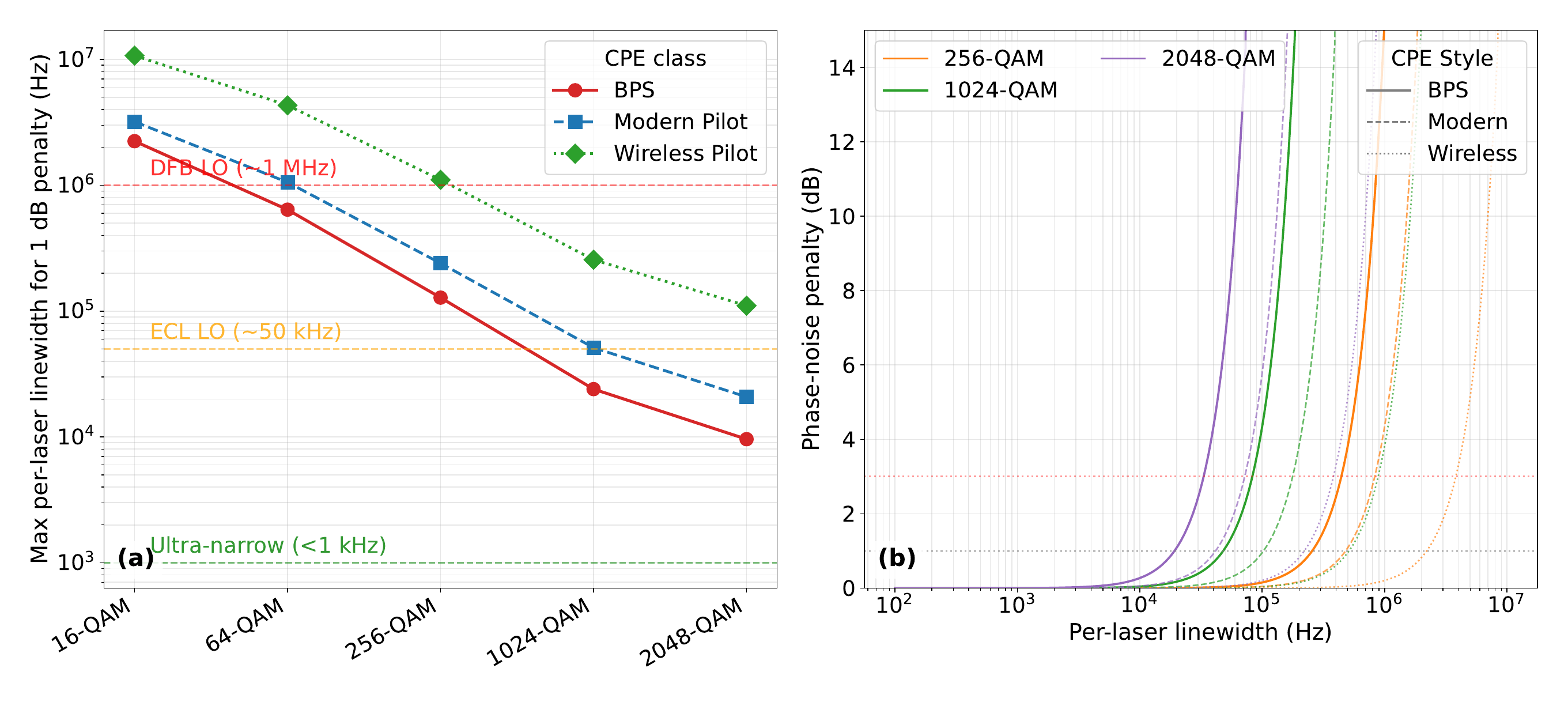}
	\vspace{-0.55cm}
	\caption{Phase-noise tolerance under the three CPE classes:
		(a)~maximum per-laser linewidth for 1~dB penalty versus QAM
		order; (b)~phase-noise penalty versus per-laser linewidth at
		high QAM orders.}
	\label{fig:phase_noise}
	\vspace{-0.55cm}
\end{figure*}

\begin{figure*}[!t]
	\centering
	\includegraphics[width=0.72\textwidth]{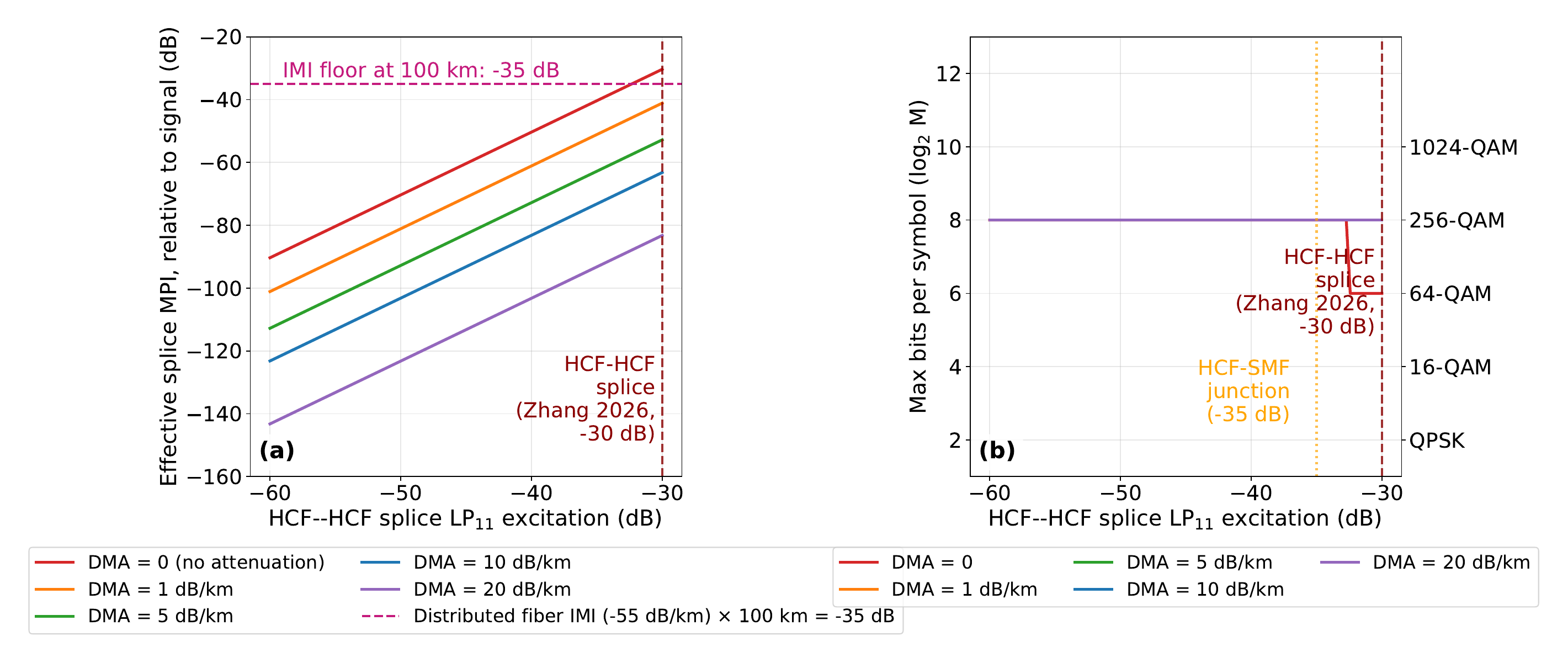}
	\vspace{-0.55cm}
	\caption{Distributed versus discrete inter-modal interference:
		(a)~effective splice MPI versus HCF--HCF splice LP$_{11}$
		excitation at several DMA values; (b)~maximum feasible QAM
		order versus splice excitation.}
	\label{fig:imi_impact}
	\vspace{-0.55cm}
\end{figure*}

\subsection{The Transceiver Ceiling Sets the Achievable Order}
\label{subsec:res_trx_ceiling}
Figure~\ref{fig:trx_ceiling} plots the transceiver back-to-back SNR
ceiling against symbol rate (Table~\ref{tab:ceiling}). For the
high-performance class the ceiling is $\approx$28.8~dB at 32~GBaud,
falling to $\approx$25.3~dB at 64~GBaud and $\approx$21.2~dB at
128~GBaud; the deployable class is $\sim$2.5--3~dB lower. Because the
1024- and 2048-QAM thresholds are $29.6$ and
$32.4$~dB, the ceiling crosses the 1024-QAM requirement near
24~GBaud and lies below the 2048-QAM requirement at essentially
all practical baud rates. The ceiling alone therefore forbids 1024-QAM
at our 64~GBaud operating point and $\ge$2048-QAM at 32~GBaud and
above, on either fiber; even 256-QAM requires the high-performance
class above $\sim$45~GBaud. The published HCF PS-1024-QAM
(20~GBaud)~\cite{Fan2026_1024QAM_DNANF} and PS-256-QAM
(60~GBaud)~\cite{Wang2024_NANF} demonstrations fall at the corresponding
boundaries of the high-performance curve, and the record SMF
$\ge$2048-QAM demonstrations, which use research-grade instrumentation
and $\le$10~GBaud, sit above our deployable-transceiver
class~\cite{Beppu2015_2048QAM,Olsson2018_4096QAM200km,Chen2019_16384QAM}.
This is the central change from the conventional OSNR-only picture and
is what reframes HCF's value as reach- and baud-at-a-given-order rather
than ultra-high QAM.

\subsection{EEPN Sensitivity}
EEPN scales with accumulated dispersion and symbol rate, so HCF's
$\sim$5.7$\times$ lower dispersion yields a proportional reduction. At
256-QAM, 1000~km and the 64~GBaud single-carrier operating point, the
EEPN penalty is $\sim$1.3~dB on SMF versus $\sim$0.25~dB on HCF
(Fig.~\ref{fig:eepn_comparison}). However, with DSCM -- the
per-subcarrier EEPN falls as $1/N_{\text{sc}}$ (each subcarrier sees a
proportionally smaller CD-equalizer group-delay spread); at
$N_{\text{sc}}=4$ this penalty drops to $\sim$0.35~dB (SMF) and
$\lesssim$0.1~dB (HCF). We model only this first-order EEPN scaling, not
a full DSCM DSP simulation. One boundary of the DSCM benefit deserves
emphasis: subcarrier multiplexing is applied digitally after the
converters, so the digital-to-analog and analog-to-digital converter (DAC/ADC) pair still processes the full channel
bandwidth at the same aggregate sample rate regardless of
$N_{\text{sc}}$. The quantization and analog-bandwidth floors of
Eq.~\eqref{eq:trx_ceiling}---and hence the B2B ceiling and the maximum
feasible QAM order at a given aggregate baud---are therefore unchanged
by DSCM; only the EEPN interaction is relaxed. EEPN is therefore a
minor term for both fibers under DSCM and ceases to be a primary
HCF differentiator at the moderate per-subcarrier baud of deployed
systems; HCF's dispersion advantage instead manifests in
high-single-carrier-baud operation and in the CD-equalizer tap count
($\propto |D|\,L\,B^2$).

\subsection{Phase-Noise Tolerance}
\label{subsec:phase_noise_results}

Figure~\ref{fig:phase_noise} plots the per-laser linewidth tolerance
under the three CPE classes. With matched
$\Delta\nu_{\text{TX}}=\Delta\nu_{\text{LO}}=50$~kHz (combined
100~kHz Wiener linewidth) the phase-noise penalty stays well below
1~dB up to 256-QAM and is approximately 1~dB at 1024-QAM under the
modern-pilot model (max-tolerance per-laser linewidth $\sim$51~kHz),
crossing 1~dB above 1024-QAM. The RCM-style model further relaxes
the per-laser linewidth budget by roughly an order of magnitude at
high QAM, comfortably accommodating 50~kHz lasers up to 1024-QAM
and tolerating much higher linewidths at 256-QAM. As emphasized in
Section~\ref{subsec:phase_noise}, the model expressions for the
modern and RCM-style improvements at $M\ge 1024$ are smooth
interpolations rather than experimentally calibrated thresholds, and
should be read as projections of HCF headroom rather than as
guaranteed operating points.

\subsection{Distributed versus Discrete IMI}
\label{subsec:res_imi}

\begin{figure*}[!t]
	\centering
	\includegraphics[width=.7\textwidth]{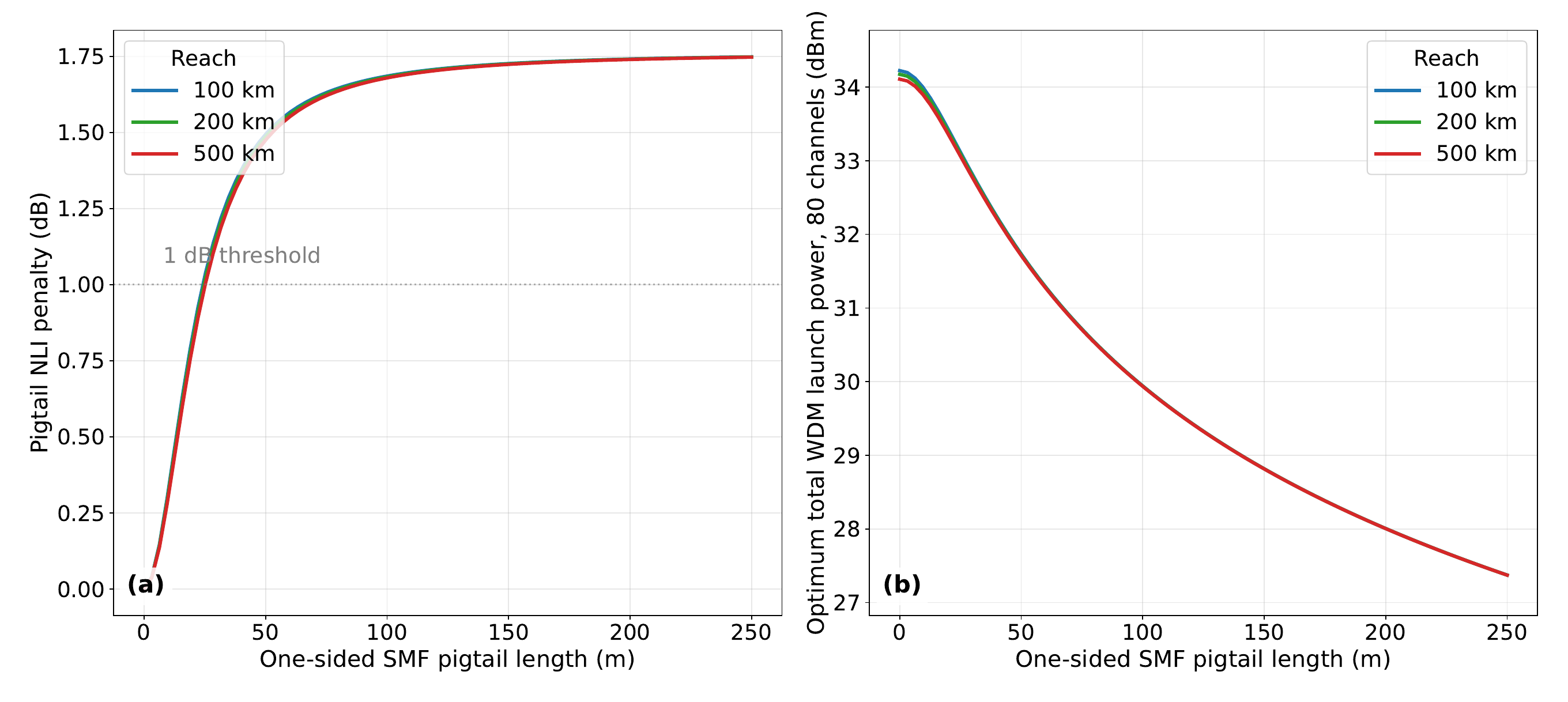}
	\vspace{-0.55cm}
	\caption{SMF-pigtail NLI in HCF coherent links:
		(a)~pigtail NLI penalty versus one-sided pigtail length; (b)~pigtail-aware
		optimum total WDM launch power versus pigtail length. The GN optimum
		stays below the $+40$~dBm high-power booster cap across the whole
		range, so the launch is set by the residual NLI, not by the cap.}
	\label{fig:pigtail_nli}
	\vspace{-0.55cm}
\end{figure*}

\begin{figure*}[!t]
	\centering
	\includegraphics[width=.63\textwidth]{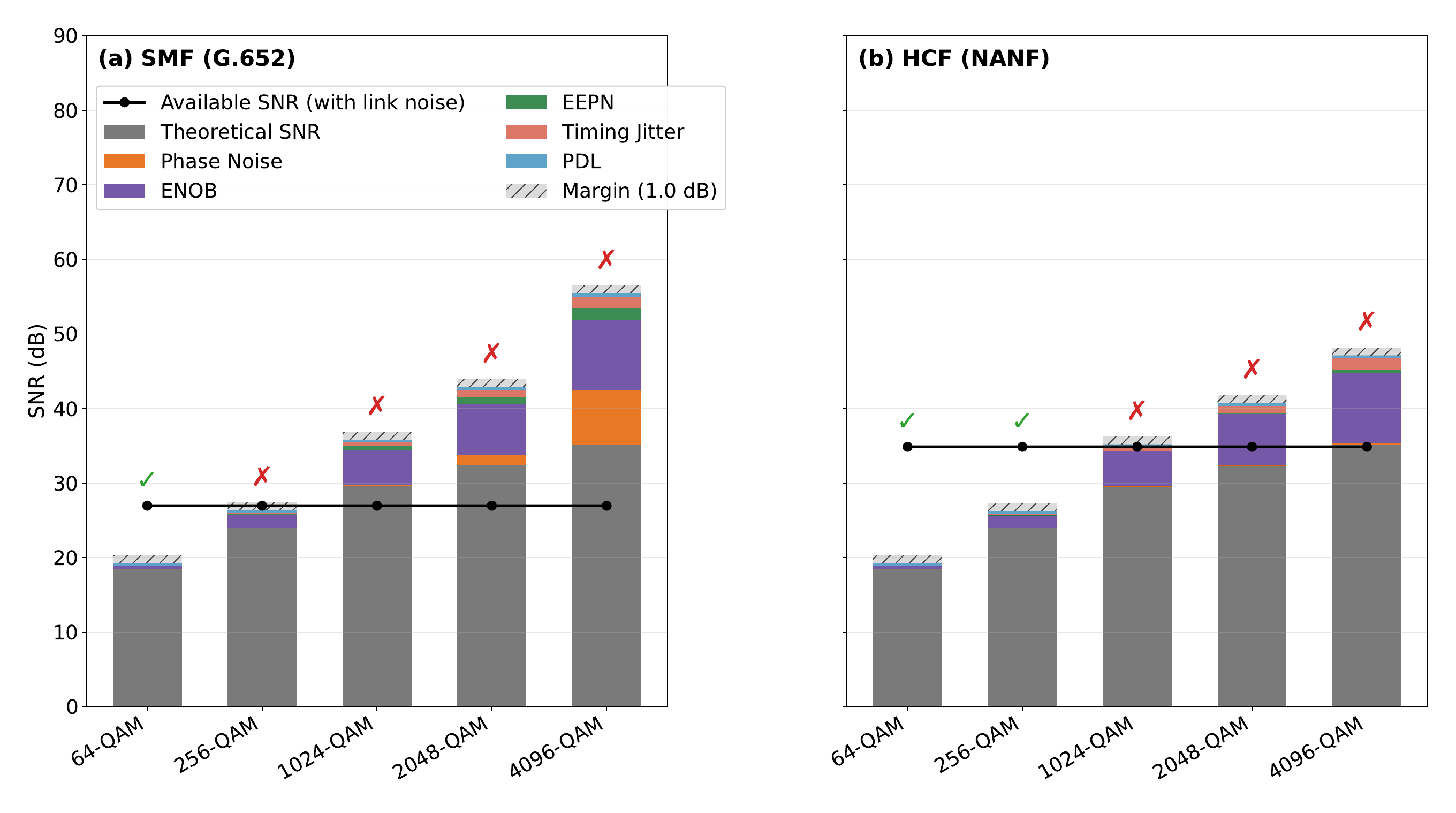}
	\vspace{-0.55cm}
	\caption{SNR penalty breakdown by QAM order at 100~km for
		(a)~SMF and (b)~HCF.}
	\label{fig:penalty_breakdown}
	\vspace{-0.55cm}
\end{figure*}
Figure~\ref{fig:imi_impact} compares discrete splice MPI to the
distributed fiber-IMI baseline. To stress-test the all-pairs MPI
accumulation, Fig.~\ref{fig:imi_impact} uses a tighter 2~km
HCF--HCF segment length, which maximizes the number of
splice/segment combinations contributing to the per-span MPI sum.
The 5~km Long-haul segment length used everywhere else in this
paper yields a strictly smaller discrete-MPI floor; the qualitative
conclusion that splice MPI is below the distributed-IMI floor at
DMA $\gtrsim 1$~dB/km is therefore conservative. The figure sweeps over the splice
LP$_{11}$ excitation in the $-60$ to $-30$~dB range---which
brackets the $\sim$$-30$~dB operating point inferred from the
fusion-splice evaluation of Zhang~\textit{et~al.}~\cite{Zhang2026}
and the $-60$~dB optimistic lab-quality target---and over the
representative DMA values 1, 5, 10 and 20~dB/km. Within this
parameter space the DMA-induced round-trip suppression keeps the
all-pairs accumulated splice MPI below the distributed-IMI floor
at $\kappa=-30$~dB excitation by a margin that grows quickly with
DMA: about 5~dB at DMA $=1$~dB/km, and an order of magnitude or
more for DMA $\gtrsim 5$~dB/km. Consequently, the maximum
feasible QAM order obtained from the SNR budget does not change
as the splice excitation is varied across this range,
including at the Zhang-inferred operating point; this is what is
plotted as a flat curve in the right panel of
Fig.~\ref{fig:imi_impact}. The result is conditional on
DMA~$\gtrsim 1$~dB/km, on the segment lengths used here, and on the
distributed-IMI coefficient $\kappa=-55$~dB/km; we do not claim it
holds at DMA~$\to 0$, where the all-pairs sum grows quadratically
with splice count and discrete splice MPI can become significant.
Within the parameter envelope studied, reducing the distributed IMI
coefficient has a larger system-level effect than improving
discrete splice excitation by typical amounts.

\subsection{SMF Pigtail Impact and Launch Power}
\label{subsec:res_pigtail}
Pigtail length and reach are independent input axes:
Fig.~\ref{fig:pigtail_nli} evaluates the pigtail-NLI penalty at fixed
reaches (100/200/500~km), with pigtail length a per-amplifier-site
property and reach setting the span count. The penalty (the dB by which
the pigtail NLI sits above the ASE floor) rises from 0~dB at
$L_{\text{pig}}=0$, crossing 1~dB near $L_{\text{pig}}\approx 25$~m and
saturating near $\sim$1.75~dB at long pigtails.
The operating launch is the smaller of the
GN-model SNR optimum (the launch that balances ASE against NLI) and the
$+40$~dBm booster cap. Across the whole studied range the GN optimum
lies below the cap, so the launch is set by the residual NLI and
not by the amplifier: even at zero pigtail HCF's own Kerr and MFA-taper
NLI cap the optimum near $\sim$$+34$~dBm total, and adding pigtail brings
it down smoothly to $\sim$$+30$~dBm total ($\sim$$+11$~dBm/ch) at 100~m
and $\sim$$+28$~dBm at 200~m (Fig.~\ref{fig:pigtail_nli}(b)). This
operating optimum sits comfortably below the $+40$~dBm cap and within the
range of aggregate launches already used in nonlinearity-free HCF
demonstrations~\cite{Hong2024_HCF_terabit,Hong2025_HCF_JLT,Sohanpal2026_launch}.
At the unconstrained optimum the NLI penalty is $\sim$1.76~dB---the
classical ASE-versus-NLI GN-model result. The $\sim$35--40~dBm launches reported
experimentally are maximum launches in nonlinearity-free
demonstrations~\cite{Hong2024_HCF_terabit,Hong2025_HCF_JLT}, where exceeding the SNR
optimum is harmless precisely because the short, low-span links there
accumulate negligible NLI. Because signal and IMI both scale linearly
with the channel power, the IMI-limited reaches reported below are
insensitive to the launch-power choice. 

\subsection{Penalty Breakdown}
\label{subsec:res_breakdown}

Figure~\ref{fig:penalty_breakdown} compares the budgets for SMF and HCF
at 100~km, 256-QAM. The HCF link SNR (Eq.~\eqref{eq:snr_link}) reaches
$\sim$34.9~dB at the $\kappa=-55$~dB/km baseline---about 7.9~dB above
the SMF link SNR of $\sim$27.0~dB---because HCF operates at higher
launch and lower loss; distributed IMI is the dominant HCF link-noise
term, pulling the SNR down by $\sim$17~dB relative to the ASE-only
floor, while discrete splice MPI is below the bar-chart resolution at
DMA~$=10$~dB/km. However, at the 64~GBaud operating point the
transceiver ceiling (Section~\ref{subsec:res_trx_ceiling}) is
$\sim$25.3~dB on the high-performance class, so the total SNR after
combination (Eq.~\eqref{eq:snr_total}) is $\sim$24.1~dB on HCF and
$\sim$22.5~dB on SMF. 256-QAM (threshold $24.0$~dB, credit $1.2$~dB) is
therefore feasible at 100~km on HCF (margin $\sim$$+0.3$~dB) but
not on SMF, where the lower link SNR caps the order at 64-QAM.
1024-QAM (threshold $29.6$~dB) is infeasible on both: the HCF link
headroom (34.9~dB) would support it, but the transceiver ceiling caps
the delivered SNR below the requirement.

Among the fiber terms the IMI coefficient is still the most
influential: at the baseline $\kappa=-55$~dB/km the distributed-IMI
floor is what terminates the 256-QAM reach at $\sim$170~km, and every
dB of $\kappa$ improvement translates directly into reach
(Fig.~\ref{fig:max_qam_distance}). Li~\textit{et~al.}~\cite{li_ofc2026}
reported measured IMI coefficients in the $-63.6$ to $-73.2$~dB/km range, including a 266~km link at $-68.8$~dB/km, so
improved-$\kappa$ regimes are being approached in laboratory fiber; the
reach leverage of such values, together with its deployed-cabling
uncertainty, is quantified in Section~\ref{subsec:res_cabling}.

\subsection{Maximum QAM Order versus Distance}

\begin{figure}[!t]
	\centering
	\includegraphics[width=0.82\columnwidth]{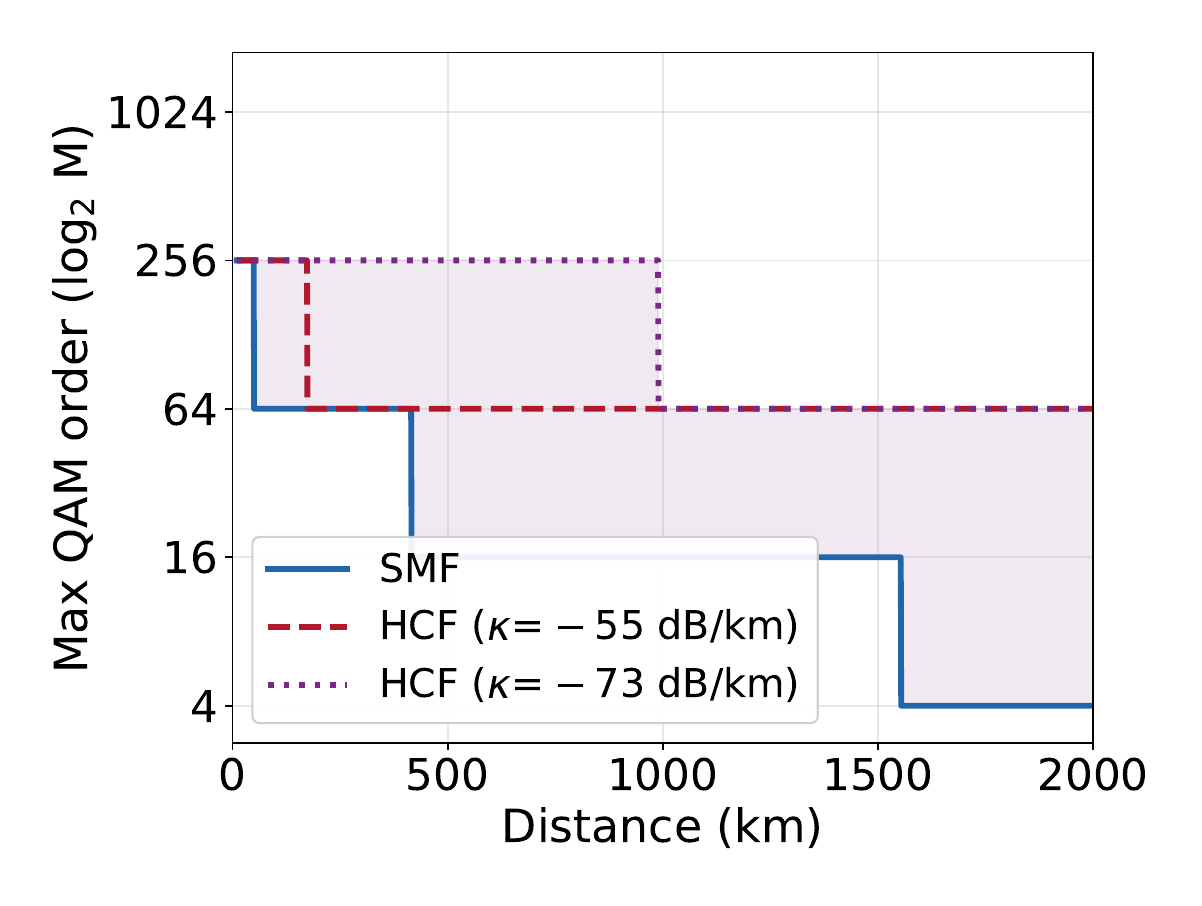}
	\vspace{-0.55cm}
	\caption{Maximum feasible QAM order versus distance for SMF and
		HCF at the baseline $\kappa=-55$~dB/km (64~GBaud,
		high-performance transceiver); the laboratory-specimen
		$\kappa=-73$~dB/km curve is included as a best-case bound only
		(see Section~\ref{subsec:res_cabling}). A gas-free core is
		assumed here (ultimate performance); the CO$_2$ gas-line
		sensitivity is treated separately in
		Section~\ref{subsec:res_gas}. At 64~GBaud the
		transceiver ceiling caps the order at 256-QAM; the curves
		separate by reach.}
	\label{fig:max_qam_distance}
	\vspace{-0.55cm}
\end{figure}
Figure~\ref{fig:max_qam_distance} shows the maximum feasible QAM order
versus distance at 64~GBaud for SMF (G.652) and HCF at the baseline
$\kappa=-55$~dB/km, {assuming} {a gas-free core (sealed/evacuated, or
equivalently line-free), so the curves show the ultimate transceiver-
and IMI-limited performance. A uniform 64~GBaud, 75~GHz grid cannot in
general place every} channel {off the CO$_2$ lines, so this assumption
isolates the fiber/transceiver limit; the CO$_2$ sensitivity is treated
separately in Section~\ref{subsec:res_gas}.} At 64~GBaud
the transceiver ceiling caps the achievable order at 256-QAM; the HCF
advantage appears as reach. At the baseline $\kappa$, the reach at each
order (5~km resolution) is:
\begin{itemize}
	\item \textbf{1024-QAM and above:} transceiver-limited and infeasible
	on either fiber at 64~GBaud; reachable only at reduced baud
	(Section~\ref{subsec:res_baudrate_heatmap}). At 16~GBaud, 1024-QAM
	reaches $\sim$85~km on baseline HCF, consistent with the
	20~GBaud, 5~km HCF PS-1024-QAM
	demonstration~\cite{Fan2026_1024QAM_DNANF}.
	\item \textbf{256-QAM:} $\sim$$45$~km on SMF versus $\sim$$170$~km on
	HCF ($\sim$$3.8\times$).
	\item \textbf{64-QAM:} $\sim$$415$~km on SMF versus $\sim$$2275$~km on
	HCF ($\sim$$5.5\times$).
	\item \textbf{16-QAM:} $\sim$1550~km on SMF and $>$2000~km on HCF.
\end{itemize}
The picture that emerges is that the HCF advantage is mediated by
the distributed IMI coefficient $\kappa$. At the conservative
$\kappa=-50$~dB/km of~\cite{Poggiolini2022_HCF} the linear-in-
$P_{\text{ch}}$ IMI floor consumes much of HCF's link headroom, while
every dB of improvement below the baseline extends the reach at each
order. The curve for the laboratory-specimen $\kappa=-73$~dB/km of
Li~\textit{et~al.}~\cite{li_ofc2026} is included in
Fig.~\ref{fig:max_qam_distance} as a best-case laboratory bound only;
because cabling and deployment are expected to degrade $\kappa$
materially, we defer quantitative reach statements for that regime to
the deployed-cabling band analysis of
Section~\ref{subsec:res_cabling}. The published HCF
demonstrations---PDM-PS-256-QAM at 60~GBaud over
NANF~\cite{Wang2024_NANF} and PDM-PS-1024-QAM at 20~GBaud over
DNANF~\cite{Fan2026_1024QAM_DNANF}---fall within the corresponding
regions of the model.

\subsection{Symbol Rate $\times$ Distance Design Space}
\label{subsec:res_baudrate_heatmap}

\begin{figure*}[!t]
	\centering
	\includegraphics[width=.68\textwidth]{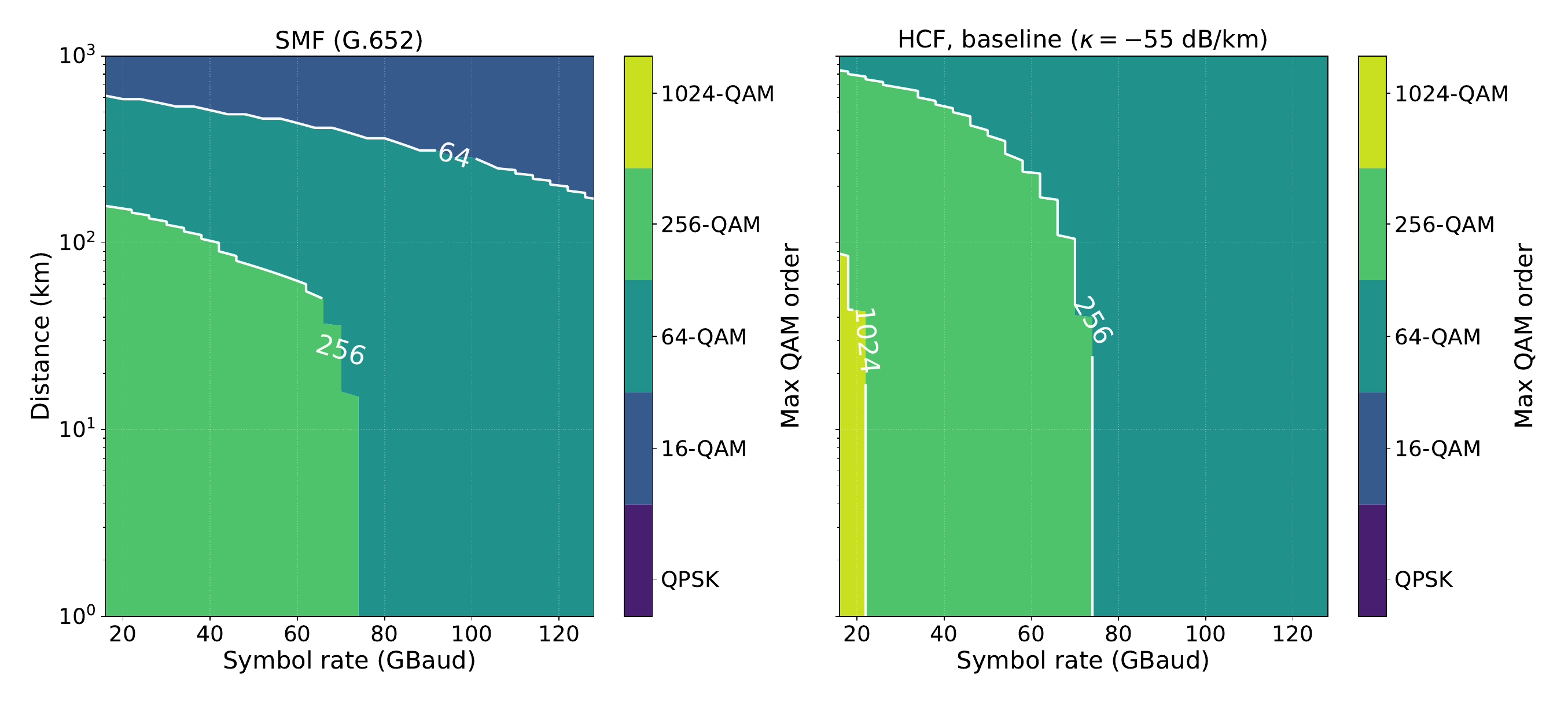}
	\vspace{-.3cm}
	\caption{Maximum feasible QAM order as a function of symbol rate
		(16--128~GBaud) and link distance for SMF (left) and HCF baseline
		(right), on a logarithmic distance axis. The color scale ends at
		1024-QAM, the highest order feasible anywhere in this range;
		higher orders require symbol rates below $\sim$10~GBaud
		(Fig.~\ref{fig:trx_ceiling}).}
	\label{fig:baudrate_distance_heatmap}
	\vspace{-.3cm}
\end{figure*}

\begin{figure*}[!t]
	\centering
	\includegraphics[width=.66\textwidth]{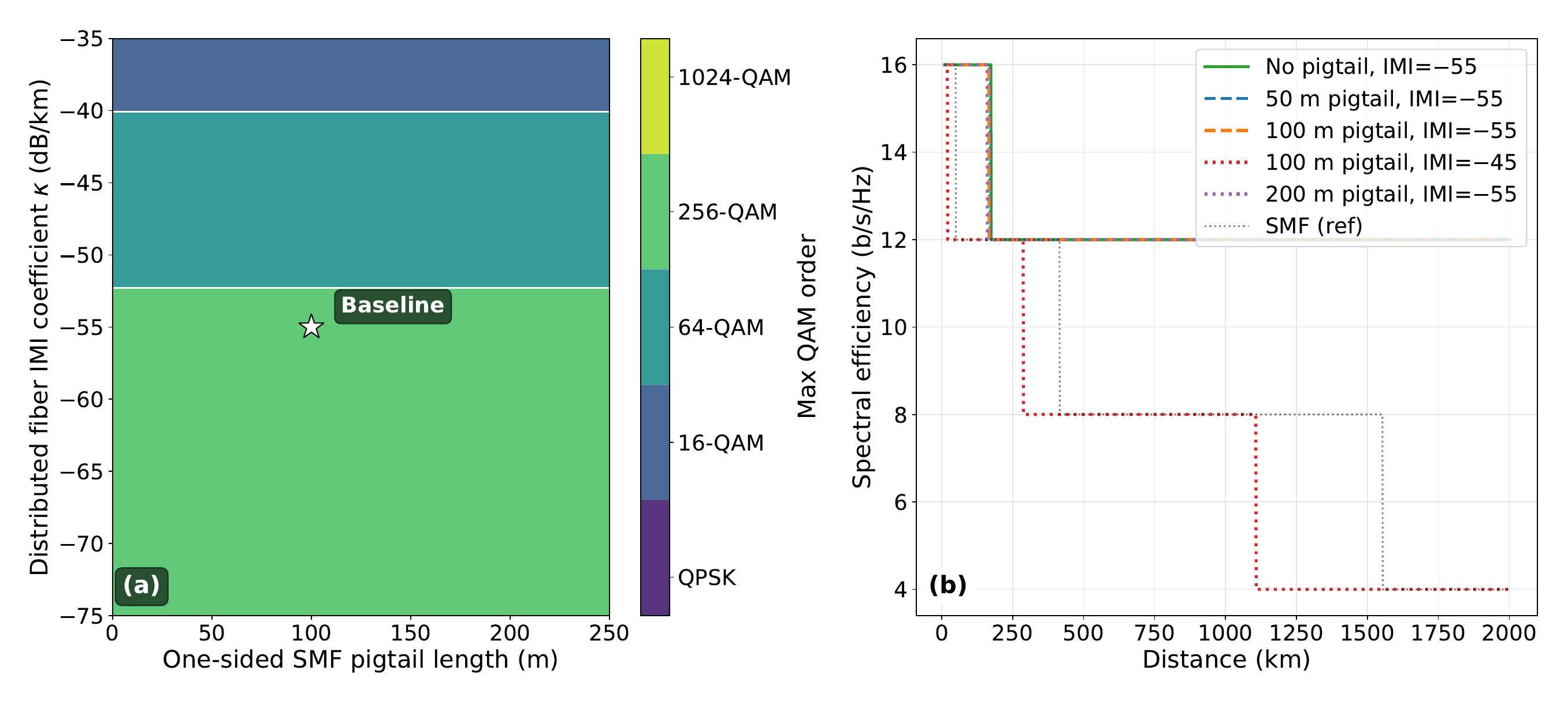}
	\vspace{-.4cm}
	\caption{(a)~Combined pigtail $\times$ distributed-IMI design
		space at 100~km; (b)~spectral efficiency versus distance for
		representative HCF operating points, with SMF as reference.}
	\label{fig:combined_design}
	\vspace{-.4cm}
\end{figure*}

Figure~\ref{fig:baudrate_distance_heatmap} maps the maximum feasible
QAM order over the joint symbol-rate $\times$ distance space for SMF and
HCF (baseline scenario) from 16~GBaud up to the 128~GBaud of
1.6T/3.2T-class transceivers~\cite{Fan2024_DSP}. The dominant feature
is the baud-vs-order trade set by the transceiver ceiling: the highest
achievable order falls monotonically with symbol rate on both fibers.
The 1024-QAM region survives only in the low-baud, short-reach corner
of the HCF panel ($\lesssim$20~GBaud, up to $\sim$85~km at 16~GBaud),
where the published ultra-high-QAM demonstrations sit; orders above
1024-QAM are infeasible everywhere in the 16--128~GBaud range, their
requirements being met only below $\sim$10~GBaud
(Fig.~\ref{fig:trx_ceiling}), so the color scale ends at 1024-QAM.
Above $\sim$32~GBaud the achievable order is 256-QAM or below
regardless of fiber---which is why deployed high-baud pluggables run
16-/64-QAM. Within each order band HCF extends the reach relative to
SMF at every symbol rate. The practical implication, in agreement with
the trend of moving to higher baud at moderate order for 1.6T/3.2T
transceivers, is that HCF's near-term value is delivered as high-baud,
moderate-order, long-reach operation rather than ultra-high QAM.

\subsection{Combined Design Space}

Figure~\ref{fig:combined_design} maps the joint pigtail $\times$
distributed-IMI design space at 100~km. With the IMI axis now
extended to span the full published range
($-35$ to $-75$~dB/km), the panel shows a striking QAM-order
staircase along $\kappa$: at the baseline operating point
(100~m pigtail, $\kappa=-55$~dB/km, marked by the white star) the
achievable order at 64~GBaud is capped at 256-QAM by the
transceiver ceiling, so within the panel the boundary is set by
$\kappa$ rather than by an order staircase: 256-QAM is supported for
$\kappa\lesssim-54$~dB/km at 100~km and degrades to 64-QAM as $\kappa$
worsens toward $-45$~dB/km, essentially independent of pigtail length
across the practical 0--200~m range. The pigtail axis is therefore a
weak parameter here---panel \textbf{(b)} shows the no-pigtail and
200-m-pigtail spectral-efficiency curves at baseline $\kappa$
overlapping almost everywhere---confirming that the binding fiber
parameter is $\kappa$, while the binding system parameter is the
transceiver ceiling. A conservative $\kappa$ envelope ($\le -50$~dB/km)
is required to preserve the HCF reach advantage at long reach.

\subsection{Gas-Line Sensitivity}
\label{subsec:res_gas}
\begin{figure*}[!t]
	\centering
	\includegraphics[width=0.83\textwidth]{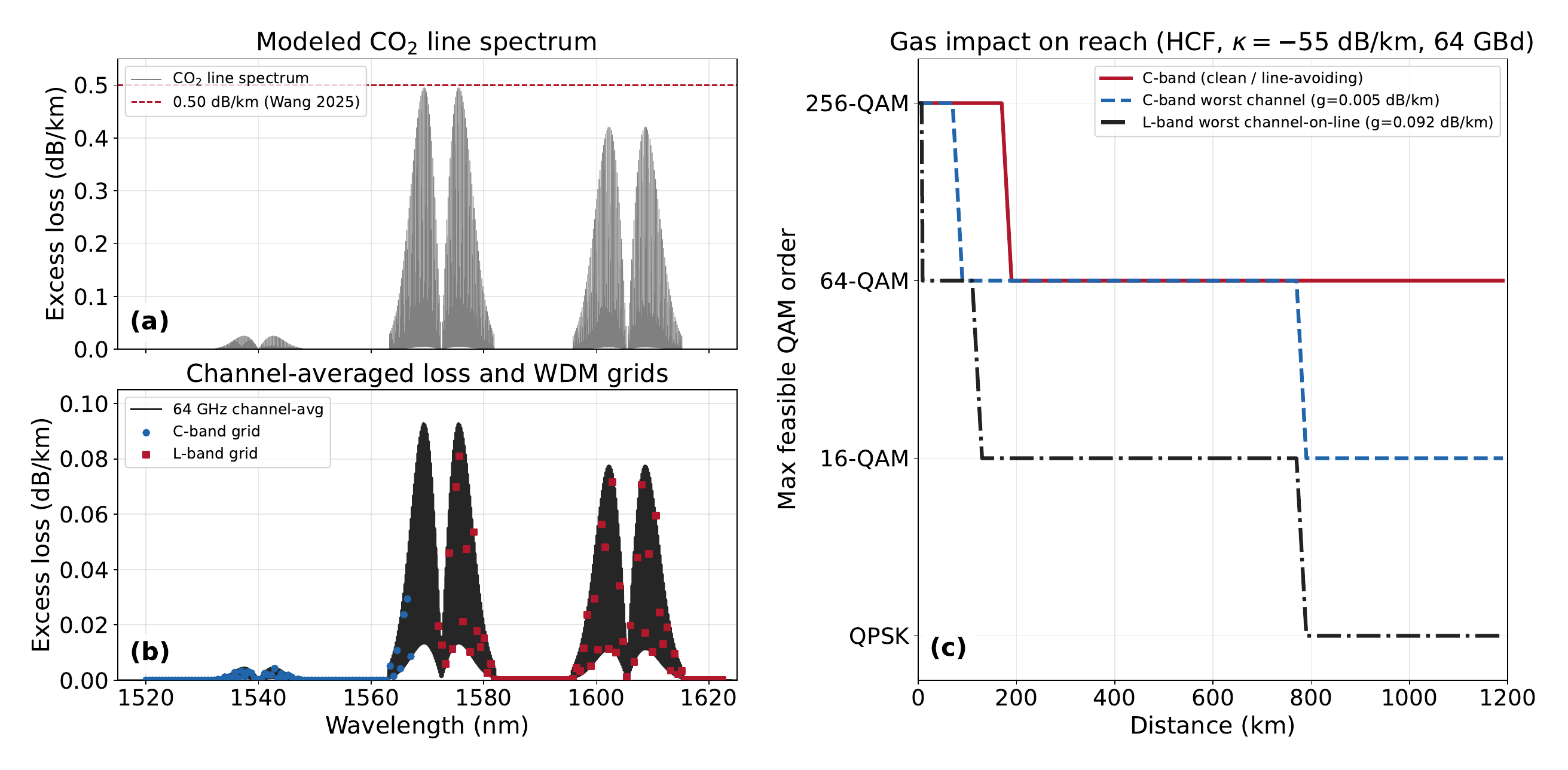}
	\vspace{-0.55cm}
	\caption{CO$_2$ gas-line absorption. (a)~Modeled CO$_2$ line
		spectrum (line-center excess loss). (b)~64~GHz channel-averaged
		excess loss with the C-band and L-band WDM grids overlaid, on a
		$\sim$5$\times$ finer vertical scale than~(a). (c)~Max feasible
		QAM order versus distance (HCF, $\kappa=-55$~dB/km, 64~GBaud) for
		a clean C-band channel, the worst C-band channel, and a worst-case
		L-band channel-on-line.}
	\label{fig:gas_sensitivity}
	\vspace{-0.55cm}
\end{figure*}

Figure~\ref{fig:gas_sensitivity}(a) shows the modeled CO$_2$ line
spectrum, and Fig.~\ref{fig:gas_sensitivity}(b) overlays the C-band grid
(1520--1567~nm) and the L-band grid (1572--1623~nm) on the 64-GHz
channel-averaged loss. {Here the \emph{channel-averaged loss} is the
per-frequency CO$_2$ excess-loss profile $g(\nu)$ of
Section~\ref{subsec:gas} averaged (power-spectral-density weighted) over
the $\sim$64~GHz channel passband; because each pressure-broadened line
is only $\sim$1.5~GHz wide, this average sits about an order of
magnitude below the line-center peak of Fig.~\ref{fig:gas_sensitivity}(a).} Operating in the C-band is essentially gas-free: the
mean per-channel excess loss is $\sim$0.0015~dB/km and only the few
reddest channels (near 1567~nm, touching the 1573~nm band edge) reach
$\sim$0.03~dB/km; the worst placement within the deep C-band
(1530--1562~nm) sees only $\sim${0.005~dB/km (channel-averaged, CO$_2$
lines only; its line-center value is $\sim$10$\times$ higher), anchored
to the up-to-0.5~dB/km line-center loss of~\cite{Wang2025_OL_CO2}.} The L-band grid, by
contrast, straddles both strong bands; about half its channels are
materially affected, the worst reaching $\sim$0.08--0.09~dB/km
(channel-averaged). Figure~\ref{fig:gas_sensitivity}(c)
translates this into reach at 64~GBaud. A representative (clean) C-band
channel follows the gas-free baseline (256-QAM to $\sim$170~km, 64-QAM to
$\sim$2275~km). The worst deep-C-band channel ($\sim$0.005~dB/km) is
indistinguishable from that baseline at metro reach, but the in-band
notch deepens with accumulated length: its 256-QAM reach shrinks to
$\sim$90~km and its 64-QAM reach is ISI-limited to $\sim$790~km, so even
the weak C-band lines become consequential once the reach is long and
the margin tight. An L-band channel-on-line fares far worse,
losing one to two QAM orders even at short reach: its 64-QAM reach
collapses from $\sim$2275~km to $\sim$120~km, and 256-QAM to a few
km. Gas absorption is thus most consequential where the
margins are tightest---high order and long reach: line-aware C-band
operation is essentially unaffected, whereas worst-case C-band
placements at long reach and all L-band channels-on-line are
CO$_2$-limited and require spectral planning, a sealed/evacuated core, or
{DSCM/pre-emphasis} mitigation~\cite{Wang2025_OL_CO2,Sillekens2026_GLA,saber2026_gla}.

\subsection{Deployed-Cabling IMI Sensitivity}
\label{subsec:res_cabling}
\begin{figure}[!t]
	\centering
	\includegraphics[width=0.92\columnwidth]{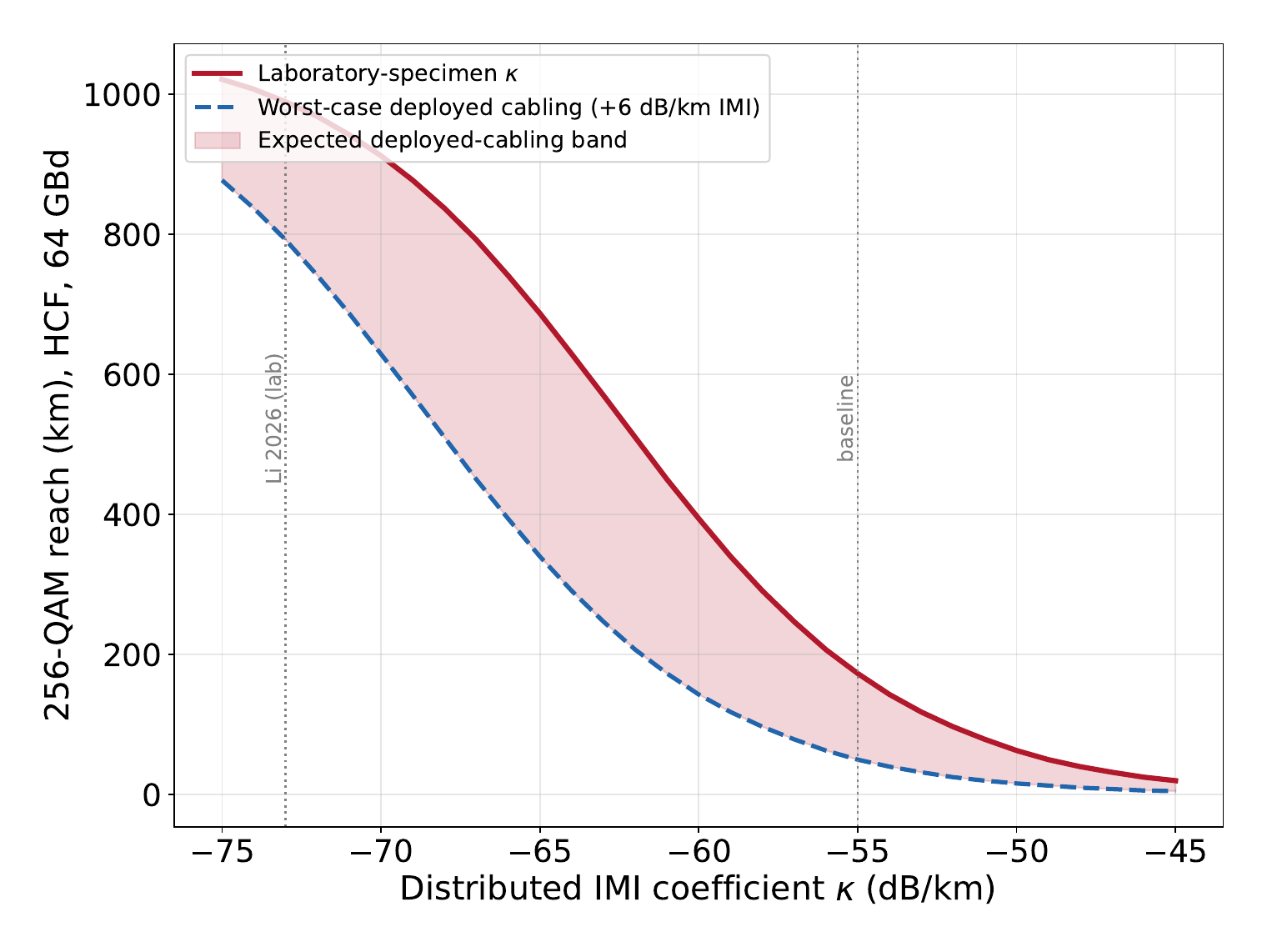}
	\vspace{-0.55cm}
	\caption{256-QAM reach versus distributed IMI coefficient $\kappa$
		(HCF, 64~GBaud) with a shaded deployed-cabling band
		(laboratory $\kappa$ to a $+6$~dB/km worst-case cabling penalty).
		Baseline $\kappa=-55$~dB/km and the laboratory Li~2026
		$\kappa=-73$~dB/km are marked.}
	\label{fig:kappa_band}
	\vspace{-0.55cm}
\end{figure}

The strong dependence of reach on $\kappa$ raises the question of how
$\kappa$ behaves in cabled, deployed fiber. Laboratory-specimen $\kappa$
is expected to be optimistic: bends, splices, mechanical stress, and
temperature variation all increase higher-order-mode content and
coupling. Published cabled-IMI BER/SNR data are scarce---a
field-deployed 0.11~dB/km HCF cable has been
demonstrated~\cite{Xiong2024} but without a cabled-$\kappa$
characterization---so we present reach as a band rather than a point.
Figure~\ref{fig:kappa_band} plots 256-QAM reach versus $\kappa$ with a
shaded band between the laboratory $\kappa$ and a worst-case deployed
penalty of $+6$~dB/km added IMI. At the baseline $\kappa=-55$~dB/km the
256-QAM reach spans $\sim$50--170~km across the band; at the
laboratory $\kappa=-73$~dB/km it spans roughly $\sim$790~km to
$\sim$975~km, the band narrowing (relatively) as the transceiver
ceiling, rather than IMI, becomes the binding constraint at low
$\kappa$. We therefore treat the most aggressive single-point figures
(e.g.\ multi-thousand-km reach at $\kappa=-73$~dB/km) as best-case
laboratory bounds, not deployment predictions, and identify cabled-IMI
BER/SNR characterization as the key open measurement for this class of
result.

\section{Discussion}
\label{sec:discussion}

\subsection{Where to Improve First}

The order of priorities changes once the transceiver ceiling is
included. At deployable symbol rates the binding constraint for
$\ge$1024-QAM is the transceiver, so the highest-leverage direction {for}
ultra-high order is transceiver improvement (higher-ENOB converters at
rate, wider analog bandwidth, lower Tx/Rx nonlinearity) or reduced baud.
For the {transceiver-feasible} orders ($\le$256-QAM at {64~GBaud)} the
{binding} fiber parameter is the distributed IMI coefficient $\kappa$:
SMF pigtails and discrete splice MPI together contribute $<\!0.5${~dB,} whereas
tightening $\kappa$ from $-55$ to $-73$~dB/km extends the 256-QAM reach
from $\sim${170} to $\sim$975~km. The support-tube HCF of
Li~\textit{et~al.}~\cite{li_ofc2026} ($-63.6$ to $-73.2$~dB/km IMI,
0.040~dB/km loss) shows this regime is being {approached;} ensuring such
low IMI survives cabling (Section~\ref{subsec:res_cabling}) is the key
fiber-side direction.

\subsection{Phase Noise and CPE Choice}

With 50~kHz ECL lasers the phase-noise penalty stays below 1~dB up to
1024-QAM under the modern-pilot class, so the RCM-style class is not
required for the headline results; it mainly {enables} cheaper
($\sim$0.5--1~MHz) lasers if HCF's quasi-linearity preserves pilot
integrity in deployment. The $G_{\text{mod}}(M)$ and $G_{\text{RCM}}(M)$
expressions are smooth {interpolations,} {unvalidated} for $M\ge 1024${.}

\subsection{Pigtail Practice}

Pigtails of practical length (0--200~m) {add} only modest NLI at the
GN-optimum launch: at 200~m the pigtail NLI sits $\sim$1.75~dB above the
ASE floor, yet the 256-QAM margin at 100~km shrinks by $<0.05$~dB because
distributed IMI and the transceiver ceiling {dominate,} so pigtail length
is not {system-limiting.} Shorter SMF tails, direct HCF amplification, or
larger-$A_{\text{eff}}$ pigtails {(}$\gamma^2\propto 1/A_{\text{eff}}^2$)
still buy margin where every dB {counts,} provided the mode-field
transitions add no excess splice loss.

\subsection{DSP Implementation Notes}
\label{subsec:dsp_constraints}

The optical-layer results are bounded in practice by DSP ASIC
implementation; the numerical statements below are order-of-magnitude
estimates consistent with the cited literature. Three points matter for
high-order QAM on HCF. First, ENOB at high sample rate tracks the Walden/
Murmann trend~\cite{Walden1999,Murmann2026,Laperle2014}: today's 400G/800G
converters give $\sim$5.5--6~bit, falling to $\sim$5.0~bit at 64~GBaud
(0.6~bit/octave), sufficient for 256-QAM~\cite{Varughese2018} but short of
the requirement for $\ge$1024-QAM at the $\gtrsim$200~GSa/s rates of
1.6~Tb/s DSPs---the converter-side origin of the transceiver ceiling.
Second, HCF's $\sim$4$\times$ lower dispersion cuts the CD-equalizer FIR
tap count by the same factor (tap count $\propto|D|LB^2$~\cite{Savory2008}),
and its quasi-linear channel lets digital-back-propagation/Volterra blocks
be scaled back, {freeing} area and power {for} higher-precision {phase}
tracking. Third, the 1024-QAM workload fits the few-watt, 3--5~nm
envelope of current pluggables~\cite{Pillai2014,Fougstedt2020,Fan2024_DSP},
so the binding constraint is converter ENOB and analog bandwidth, not
gate-count or power.

\subsection{Consistency with Published Experiments}
\label{subsec:expt_survey}

{These} comparisons are consistency {checks,} {not} a validation of the
$\kappa$-dependent model: each demonstrated point lies inside the
model's feasible region at its reported parameters. The published HCF
{high-order} {ceiling---PDM-PS-256-QAM} at 60~GBaud over 2~km
NANF~\cite{Wang2024_NANF} and PDM-PS-1024-QAM at 20~GBaud over 5~km
DNANF~\cite{Fan2026_1024QAM_DNANF}{---sits} at the corresponding
{transceiver-ceiling} boundaries (Fig.~\ref{fig:trx_ceiling}, margins
$\sim$$+1.0$ and $\sim$$+0.5${~dB), while capacity-oriented} HCF
experiments over $>10\,000$~km {of}
AR-HCF~\cite{Ge2025_AR_HCF,Hong2024_HCF_terabit} fall in the {low-order,}
long-reach {region.} {The} {SMF} $\ge$2048-QAM
{records~\cite{Chen2019_16384QAM,Olsson2018_4096QAM200km,Beppu2015_2048QAM,Koizumi2012_1024QAM}}
all use $\le$10--30~GBaud, {shaping,} and research-grade {instrumentation,}
which the model places above its {deployable} class and only inside the
high-performance class at low baud---consistent with the absence of any
deployable $\ge$1024-QAM point at 64~GBaud.

\subsection{Model Scope and Limitations}
\label{subsec:limitations}
Table~\ref{tab:limitations} summarizes the principal approximations,
their estimated magnitude at 256- and 1024-QAM, and the direction of
the bias. The two most important residuals are: (i)~the line-system
penalty, which the baseline omits (point-to-point amplified line); a
worked traversal of three ROADMs adds $\sim$1.1~dB at 256-QAM and
$\sim$1.2~dB at 1024-QAM, an optimistic omission for ROADM-rich paths. (ii)~the gas model uses
a structured CO$_2$ line comb with a two-part penalty---excess
loss$\rightarrow$ASE (largely masked on the IMI-limited HCF link) plus a
bounded notch-ISI penalty anchored to Sillekens and
Sohanpal~\cite{Sillekens2026_GLA}; the ISI saturation level
$\delta_{\max}$ is the main calibration freedom, and a full split-step
DSP simulation of the notch is out-of-scope for the current study. Other approximations---the smooth
$G_{\text{PCS}}(M)$ interpolation, the estimated MFA-taper coefficient,
and the frequency-flat aperture-jitter model at very high baud---are
sub-dB in the operating regime and are noted where they appear. The
transceiver-ceiling parameters are anchored to reported converter and
front-end trends; their absolute calibration shifts the exact
crossover baud but not the qualitative conclusion that high-order QAM
at deployable rates is transceiver-limited.

\begin{table}[!t]
	\caption{Model Scope and Limitations}
	\label{tab:limitations}
	\centering
	\footnotesize
	\begin{tabular}{p{2.5cm}cp{1.3cm}p{1.6cm}}
		\toprule
		\textbf{Effect} & \textbf{Modeled?} & \textbf{Est. @256/1024} & \textbf{Bias} \\
		\midrule
		ROADM/WSS filtering & optional & 0/1.1--1.2~dB & optimistic \\
		Gas ASE part (HCF) & yes & masked by IMI & --- \\
		Gas notch ISI & yes (bounded) & 0/L-band only & calib.\ $\delta_{\max}$ \\
		Gas full split-step & out-of-scope & --- & --- \\
		MFA taper NLI & yes (est.) & $<$0.3~dB & conservative \\
		Aperture jitter (HF) & flat & $<$0.2~dB & optimistic \\
		$G_{\text{PCS}}(M)$ interp. & yes & $\pm$0.3~dB & --- \\
		TRx absolute calib. & anchored & shifts baud & --- \\
		\bottomrule
	\end{tabular}
\end{table}

\section{Conclusion}
\label{sec:conclusion}
A per-channel effective-SNR budget that unifies the HCF-specific fiber
terms with a rate-dependent transceiver ceiling overturns the OSNR-only
expectation: at deployable symbol rates the transceiver back-to-back
limit, not the fiber, sets the achievable QAM order. The practical
consequence is a reorientation of where HCF pays off. Ultra-high-order
QAM occupies a low-baud, short-reach corner that no fiber improvement
can enlarge, so HCF's near-term value lies instead in moderate-order,
high-baud, long-reach operation and in the headroom it frees in the
laser-linewidth and DSP-complexity budgets. Converting the low-IMI
reach promise into deployed links now hinges on cabled-$\kappa$
characterization rather than on further loss reduction, whereas raising
the achievable order awaits converter and analog-bandwidth progress in
the transceiver itself. Line-aware C-band planning renders CO$_2$
absorption negligible, while L-band operation requires a sealed core or DSP-based mitigation strategies.
{These limits are analytical---derived from a component-anchored
effective-SNR budget and cross-checked against published HCF and SMF
demonstrations rather than a dedicated experiment---so direct
experimental verification at deployable baud remains important future
work.}
Taken together, these results indicate that HCF's advantage is real
but far narrower than low-loss, low-nonlinearity intuition suggests, and
the constraint that binds is the transceiver.

\section*{Disclaimer and Acknowledgment}

The views and opinions expressed in this document belong solely to the author and do not reflect Huawei's official stance. Generative AI was used for language and grammar enhancements.

\ifCLASSOPTIONcaptionsoff
  \newpage
\fi

\bibliographystyle{IEEEtran}
\bibliography{corrected_JLT_v3}

@misc{saber2026protectionswitchinghybridhollowcore,
      title={Protection Switching in Hybrid Hollow-Core and Single-Mode Fiber Networks: Challenges, Analysis, and Mitigation Strategies}, 
      author={Md Ghulam Saber and Zhiping Jiang},
      year={2026},
      eprint={2606.23554},
      archivePrefix={arXiv},
      primaryClass={cs.NI},
      url={https://arxiv.org/abs/2606.23554}, 
}

@inproceedings{saber2026ecoc,
  author    = {Md Ghulam Saber and Zhiping Jiang},
  title     = {Hollow-Core and Standard Single-Mode Fiber Hybrid Optical Networks: A Multi-Topology Investigation of Protection Switching},
  booktitle = {The 52nd European Conference on Optical Communication (ECOC 2026)},
  address   = {Malaga, Spain},
  month     = {September},
  year      = {2026}
}

@Article{saber2026_gla,
AUTHOR = {Saber, Md Ghulam and Jiang, Zhiping},
TITLE = {{CO2}-Limited Hollow-Core Fiber Links: A Capacity-Map Guide to Pre-Emphasis and Spectral Avoidance},
JOURNAL = {Photonics},
VOLUME = {13},
YEAR = {2026},
NUMBER = {6},
ARTICLE-NUMBER = {559},
URL = {https://www.mdpi.com/2304-6732/13/6/559},
ISSN = {2304-6732},
DOI = {10.3390/photonics13060559}
}

@Article{saber2026_perspective,
AUTHOR = {Saber, Md Ghulam and Jiang, Zhiping},
TITLE = {Beyond Silica Assumptions: Optical Network Design in the Hollow-Core Era},
JOURNAL = {Photonics},
VOLUME = {13},
YEAR = {2026},
NUMBER = {7},
ARTICLE-NUMBER = {670},
URL = {https://www.mdpi.com/2304-6732/13/7/670},
ISSN = {2304-6732},
DOI = {10.3390/photonics13070670}
}

@article{Wang2025_OL_CO2,
  author  = {Xishuo Wang and Zhipei Li and Lipeng Feng and Peng Li and Kai Lv and Xia Sheng and Yuyang Liu and Ran Gao and Xiangjun Xin and Lei Zhang and Jie Luo and Anxu Zhang and Xiaoli Huo and Tao Ma and Zhenfang Wang},
  title   = {Simulation investigation of {CO\textsubscript{2}} absorption-induced performance degradation in hollow-core fiber transmission systems and spectrum-optimized digital sub-carrier multiplexing design for circumvention},
  journal = {Optics Letters},
  volume  = {50},
  number  = {22},
  pages   = {7171--7174},
  year    = {2025},
  doi     = {10.1364/OL.577505}
}

@inproceedings{Sillekens2026_GLA,
  author    = {Eric Sillekens and Ronit Sohanpal},
  title     = {Gas Line Absorption Mitigation in Hollow-Core Fibre using
               Spectral Pre-Equalisation},
  booktitle = {Proceedings of Optical Fiber Communication Conference (OFC)},
  year      = {2026},
  pages     = {Th2A.50}
}

@inproceedings{Yoon2000BEP_MQAM,
  author    = {Dongweon Yoon and Kyongkuk Cho and Jinsock Lee},
  title     = {Bit Error Probability of {M}-ary Quadrature Amplitude Modulation},
  booktitle = {IEEE Vehicular Technology Conference (VTC 2000-Fall)},
  volume    = {5},
  pages     = {2422--2427},
  year      = {2000},
  doi       = {10.1109/VETECF.2000.883298}
}

@ARTICLE{saber2026_hcfimdd,
      title={Hollow-Core Fiber in Direct-Detection Optical Networks: Technology Readiness, Deployment Drivers, and Adoption Outlook}, 
      author={Md Ghulam Saber and Zhiping Jiang},
 journal={IEEE Network}, 
      year={2026},
      volume={Early Access},
  number={},
  doi={10.1109/MNET.2026.3712562  }}

@article{Hahn2026_LPM,
  author  = {Choloong Hahn and Junho Chang and Qingyi Guo and Zhiping Jiang},
  title   = {Longitudinal Performance Monitoring Towards Hollow Core Fiber
             Systems via Node Nonlinearity},
  journal = {IEEE Photonics Technology Letters},
  volume  = {38},
  number  = {2},
  pages   = {105--108},
  year    = {2026},
  doi     = {10.1109/LPT.2025.3618638}
}

@inproceedings{Hahn2026_OFC,
  author    = {Choloong Hahn and Yunfan Xu and Xianchao Guan and Luo Han
               and Zhiping Jiang},
  title     = {Extending Longitudinal Performance Monitoring to Hollow Core
               Fiber Systems: Feasibility and Implementation},
  booktitle = {Proceedings of Optical Fiber Communication Conference (OFC)},
  year      = {2026},
  pages     = {M4K.1}
}

@article{Wang2024_NANF,
  author  = {C. Wang and K. Wang and J. Long and W. Zhou and F. Zhao
             and L. Shen and P. Li and J. Yu},
  title   = {High-Order {QAM} {NANF} Transmission Utilizing {MIMO}
             Equalizer Integrated with Low-Complexity
             Decision-Directed Carrier Phase Estimation},
  journal = {Optics Letters},
  volume  = {49},
  number  = {9},
  pages   = {2293--2296},
  year    = {2024},
  doi     = {10.1364/OL.519042},

}

@inproceedings{li_ofc2026,
  author    = {P. Li and G. Chen and A. Jia and H. Li and J. Chu
               and Y. Liu and L. Zhang and L. Zhang and J. Luo},
  title     = {Low Intermodal Interference and Low Loss Hollow Core Fibers},
  booktitle = {Proceedings of Optical Fiber Communication Conference (OFC)},
  year      = {2026},
  pages     = {M2J.1},
}

@article{Poletti2014,
  author  = {F. Poletti},
  title   = {Nested Antiresonant Nodeless Hollow Core Fiber},
  journal = {Optics Express},
  volume  = {22},
  number  = {20},
  pages   = {23807--23828},
  year    = {2014},
  doi     = {10.1364/OE.22.023807}
}

@inproceedings{Jasion2022,
  author    = {G. T. Jasion and H. Sakr and J. R. Hayes
               and S. R. Sandoghchi and L. Hooper and E. {Numkam Fokoua}
               and A. Saljoghei and H. C. H. Mulvad and M. Alonso
               and A. Taranta and T. D. Bradley and I. A. Davidson
               and Y. Chen and D. J. Richardson and F. Poletti},
  title     = {0.174\,{dB/km} Hollow Core Double Nested Antiresonant Nodeless
               Fiber ({DNANF})},
  booktitle = {Proceedings of Optical Fiber Communication Conference (OFC)},
  year      = {2022},
  pages     = {Th4C.7},
  doi       = {10.1364/OFC.2022.Th4C.7}
}

@article{Petrovich2025,
  author  = {M. Petrovich and E. {Numkam Fokoua} and Y. Chen and H. Sakr
             and A. I. Adamu and R. Hassan and D. Wu and R. {Fatobene Ando}
             and A. Papadimopoulos and S. R. Sandoghchi and G. Jasion and F. Poletti},
  title   = {Broadband Optical Fibre with an Attenuation Lower than 0.1
             Decibel per Kilometre},
  journal = {Nature Photonics},
  volume  = {19},
  pages   = {1203--1208},
  year    = {2025},
  doi     = {10.1038/s41566-025-01747-5}
}

@article{Poggiolini2022_HCF,
  author  = {P. Poggiolini and F. Poletti},
  title   = {Opportunities and Challenges for Long-Distance Transmission in
             Hollow-Core Fibres},
  journal = {Journal of Lightwave Technology},
  volume  = {40},
  number  = {6},
  pages   = {1605--1616},
  year    = {2022},
  doi     = {10.1109/JLT.2021.3140114}
}

@article{Fokoua2023,
  author  = {E. {Numkam Fokoua} and S. {Abokhamis Mousavi} and G. T. Jasion
             and D. J. Richardson and F. Poletti},
  title   = {Loss in Hollow-Core Optical Fibers: Mechanisms, Scaling Rules, and Limits},
  journal = {Advances in Optics and Photonics},
  volume  = {15},
  number  = {1},
  pages   = {1--85},
  year    = {2023},
  doi     = {10.1364/AOP.470592}
}

@article{Nespola2021,
  author  = {A. Nespola and S. Straullu and T. D. Bradley and K. Harrington
             and H. Sakr and G. T. Jasion and E. {Numkam Fokoua} and Y. Jung
             and Y. Chen and J. R. Hayes and F. Forghieri and D. J. Richardson
             and F. Poletti and G. Bosco and P. Poggiolini},
  title   = {Transmission of 61 {C}-Band Channels Over Record Distance of
             Hollow-Core-Fiber With {L}-Band Interferers},
  journal = {Journal of Lightwave Technology},
  volume  = {39},
  number  = {3},
  pages   = {813--820},
  year    = {2021},
  doi     = {10.1109/JLT.2020.3047670}
}

@book{Proakis2008,
  author    = {J. G. Proakis and M. Salehi},
  title     = {Digital Communications},
  edition   = {5th},
  publisher = {McGraw-Hill},
  address   = {New York, NY, USA},
  year      = {2008}
}

@article{Pfau2009,
  author  = {T. Pfau and S. Hoffmann and R. No\'{e}},
  title   = {Hardware-Efficient Coherent Digital Receiver Concept With
             Feedforward Carrier Recovery for {$M$-QAM} Constellations},
  journal = {Journal of Lightwave Technology},
  volume  = {27},
  number  = {8},
  pages   = {989--999},
  year    = {2009},
  doi     = {10.1109/JLT.2008.2010511}
}

@article{Sasai2020,
  author  = {T. Sasai and A. Matsushita and M. Nakamura and S. Okamoto
             and F. Hamaoka and Y. Kisaka},
  title   = {Laser Phase Noise Tolerance of Uniform and Probabilistically
             Shaped {QAM} Signals for High Spectral Efficiency Systems},
  journal = {Journal of Lightwave Technology},
  volume  = {38},
  number  = {2},
  pages   = {439--446},
  year    = {2020},
  doi     = {10.1109/JLT.2019.2945470}
}

@article{Mazur2019,
  author  = {Mikael Mazur and Jochen Schr\"{o}der and Abel Lorences-Riesgo
             and Tsuyoshi Yoshida and Magnus Karlsson and Peter A. Andrekson},
  title   = {Overhead-Optimization of Pilot-Based Digital Signal Processing
             for Flexible High Spectral Efficiency Transmission},
  journal = {Optics Express},
  volume  = {27},
  number  = {17},
  pages   = {24654--24669},
  year    = {2019},
  doi     = {10.1364/OE.27.024654}
}

@article{Fang2024,
  author  = {Xiansong Fang and Yixiao Zhu and Xiang Cai and Weisheng Hu
             and Zhixue He and Shaohua Yu and Fan Zhang},
  title   = {Overcoming Laser Phase Noise for Low-Cost Coherent Optical Communication},
  journal = {Nature Communications},
  volume  = {15},
  pages   = {6339},
  year    = {2024},
  doi     = {10.1038/s41467-024-50439-1}
}

@article{Shi2024_pilot,
  author  = {X. Shi and M. Gao and X. Huang and J. Fan and X. Han
             and X. You and G. Shen},
  title   = {Optimized Pilot Structure for {PS-PDM} Ultrahigh-Order {QAM}
             Coherent Optical Transmission},
  journal = {Optics Letters},
  volume  = {49},
  number  = {6},
  pages   = {1579--1582},
  year    = {2024},
  doi     = {10.1364/OL.519424}
}

@article{Walden1999,
  author  = {R. H. Walden},
  title   = {Analog-to-Digital Converter Survey and Analysis},
  journal = {IEEE Journal on Selected Areas in Communications},
  volume  = {17},
  number  = {4},
  pages   = {539--550},
  year    = {1999},
  doi     = {10.1109/49.761034}
}

@article{Varughese2018,
  author  = {S. Varughese and J. Langston and V. A. Thomas
             and S. Tibuleac and S. E. Ralph},
  title   = {Frequency Dependent {ENoB} Requirements for {$M$-QAM} Optical
             Links: An Analysis Using an Improved Digital to Analog
             Converter Model},
  journal = {Journal of Lightwave Technology},
  volume  = {36},
  number  = {18},
  pages   = {4082--4089},
  year    = {2018},
  doi     = {10.1109/JLT.2018.2859637}
}

@misc{Murmann2026,
  author       = {B. Murmann},
  title        = {{ADC} Performance Survey 1997--2026},
  howpublished = {[Online]. Available: \url{https://github.com/bmurmann/ADC-survey}},
  year         = {2026}
}

@article{Sun2020_DSCM,
  author  = {Han Sun and Mehdi Torbatian and Mehdi Karimi and Robert Maher
             and Sandy Thomson and others},
  title   = {{800G DSP ASIC} Design Using Probabilistic Shaping and Digital
             Sub-Carrier Multiplexing},
  journal = {Journal of Lightwave Technology},
  volume  = {38},
  number  = {17},
  pages   = {4744--4756},
  year    = {2020},
}

@article{Shieh2008,
  author  = {W. Shieh and K.-P. Ho},
  title   = {Equalization-Enhanced Phase Noise for Coherent-Detection Systems
             Using Electronic Digital Signal Processing},
  journal = {Optics Express},
  volume  = {16},
  number  = {20},
  pages   = {15718--15727},
  year    = {2008},
  doi     = {10.1364/OE.16.015718}
}

@article{Poggiolini2012,
  author  = {P. Poggiolini},
  title   = {The {GN} Model of Non-Linear Propagation in Uncompensated
             Coherent Optical Systems},
  journal = {Journal of Lightwave Technology},
  volume  = {30},
  number  = {24},
  pages   = {3857--3879},
  year    = {2012},
  doi     = {10.1109/JLT.2012.2217729}
}

@article{Poggiolini2015_eGN,
  author  = {P. Poggiolini and G. Bosco and A. Carena and V. Curri
             and Y. Jiang and F. Forghieri},
  title   = {A Simple and Effective Closed-Form {GN} Model Correction
             Formula Accounting for Signal Non-Gaussian Distribution},
  journal = {Journal of Lightwave Technology},
  volume  = {33},
  number  = {2},
  pages   = {459--473},
  year    = {2015},
  doi     = {10.1109/JLT.2014.2387891}
}

@article{Suslov2021,
  author  = {Dmytro Suslov and Mat\v{e}j Komanec and Eric R. {Numkam Fokoua}
             and Daniel Dousek and Ailing Zhong and Stanislav Zv\'{a}novec
             and Thomas D. Bradley and Francesco Poletti and David J. Richardson
             and Radan Slav\'{i}k},
  title   = {Low Loss and High Performance Interconnection Between Standard
             Single-Mode Fiber and Antiresonant Hollow-Core Fiber},
  journal = {Scientific Reports},
  volume  = {11},
  pages   = {8799},
  year    = {2021},
  doi     = {10.1038/s41598-021-88065-2}
}

@article{Fan2026_1024QAM_DNANF,
  author  = {Jiamin Fan and Yu Qin and Tingyu Fu and Jie Zhu and Yichun Shen and Limin Xiao and Mingyi Gao},
  title   = {Hierarchical {MIMO equalizer for PDM PS-1024-QAM coherent optical transmission over DNANF}},
  journal = {Optics Communications},
  volume  = {607},
  pages   = {132941},
  year    = {2026},
  doi     = {10.1016/j.optcom.2026.132941}
}

@article{Shi2024_connector,
  author  = {B. Shi and C. Zhang and T. Kelly and X. Wei and M. Ding
             and M. Huang and S. Fu and F. Poletti and R. Slav\'{i}k},
  title   = {Splicing Hollow-Core Fiber with Standard Glass-Core Fiber
             with Ultralow Back-Reflection and Low Coupling Loss},
  journal = {ACS Photonics},
  volume  = {11},
  number  = {8},
  pages   = {3288--3295},
  year    = {2024},
  doi     = {10.1021/acsphotonics.4c00677}
}

@article{Suslov2022_backref,
  author  = {Dmytro Suslov and Eric {Numkam Fokoua} and Daniel Dousek
             and Ailing Zhong and Stanislav Zv\'{a}novec
             and Thomas D. Bradley and Francesco Poletti
             and David J. Richardson and Mat\v{e}j Komanec and Radan Slav\'{i}k},
  title   = {Low Loss and Broadband Low Back-Reflection Interconnection
             Between a Hollow-Core and Standard Single-Mode Fiber},
  journal = {Optics Express},
  volume  = {30},
  number  = {20},
  pages   = {37006--37014},
  year    = {2022},
  doi     = {10.1364/OE.460635}
}

@article{Zhang2026,
  title={Fusion Splicing Performance Evaluation of Hollow-Core Fiber to Hollow-Core Fiber With Different Structural Parameters},
  author={Zhang, Cong and Zhen, Yu and Wu, Xinghuan and Li, Jianping and Qin, Yuwen and Fu, Songnian},
  journal={Journal of Lightwave Technology},
  volume={44},
  number={2},
  pages={659--664},
  year={2026},
  publisher={IEEE}
}

@article{Pillai2014,
  author  = {B. S. G. Pillai and B. Sedighi and K. Guan and N. P. Anthapadmanabhan
             and W. Shieh and K. J. Hinton and R. S. Tucker},
  title   = {End-to-End Energy Modeling and Analysis of Long-Haul Coherent
             Transmission Systems},
  journal = {Journal of Lightwave Technology},
  volume  = {32},
  number  = {18},
  pages   = {3093--3111},
  year    = {2014},
  doi     = {10.1109/JLT.2014.2331086}
}

@article{Laperle2014,
  author  = {C. Laperle and M. O'Sullivan},
  title   = {Advances in High-Speed {DACs}, {ADCs}, and {DSP} for Optical
             Coherent Transceivers},
  journal = {Journal of Lightwave Technology},
  volume  = {32},
  number  = {4},
  pages   = {629--643},
  year    = {2014},
  doi     = {10.1109/JLT.2013.2284134}
}

@inproceedings{Fan2024_DSP,
  author    = {S. H. Fan and R. L. Nguyen and J. L. C. Lust
               and H. Chien and S. Wang},
  title     = {Toward 1.6{T} Low-Power Coherent {DSP}: Challenges, and Lessons
               Learned from Preceding Generations},
  booktitle = {Proceedings of Optical Fiber Communication Conference (OFC)},
  year      = {2024},
  pages     = {M2H.1},
}

@inproceedings{Fougstedt2020,
  author    = {C. Fougstedt and O. Gustafsson and C. Bae and E. B\"{o}rjeson
               and P. Larsson-Edefors},
  title     = {{ASIC} Design Exploration for {DSP} and {FEC} of
               400-{Gbit/s} Coherent Data-Center Interconnect Receivers},
  booktitle = {Proceedings of Optical Fiber Communication Conference (OFC)},
  year      = {2020},
  pages     = {Th2A.38}
}

@misc{OIF400ZR,
  author       = {{Optical Internetworking Forum}},
  title        = {{OIF}-400{ZR}-02.0 -- Implementation Agreement
                  for {400ZR}},
  year         = {2022},
  note         = {[Online]. Available:
                  \url{https://www.oiforum.com/wp-content/uploads/OIF-400ZR-02.0.pdf}}
}

@misc{OIF800ZR,
  author       = {{Optical Internetworking Forum}},
  title        = {{OIF}-800{ZR}-01.0 -- Implementation Agreement
                  for {800ZR}},
  year         = {2024},
  note         = {[Online]. Available:
                  \url{https://www.oiforum.com/wp-content/uploads/OIF-800ZR-01.0.pdf}}
}

@article{ChoWinzer2019,
  author  = {J. Cho and P. J. Winzer},
  title   = {Probabilistic Constellation Shaping for Optical Fiber
             Communications},
  journal = {Journal of Lightwave Technology},
  volume  = {37},
  number  = {6},
  pages   = {1590--1607},
  year    = {2019},
  doi     = {10.1109/JLT.2019.2898855}
}

@article{Fehenberger2016,
  author  = {Tobias Fehenberger and Domani\v{c} Lavery and Robert Maher
             and Alex Alvarado and Polina Bayvel and Norbert Hanik},
  title   = {Sensitivity Gains by Mismatched Probabilistic Shaping for
             Optical Communication Systems},
  journal = {IEEE Photonics Technology Letters},
  volume  = {28},
  number  = {7},
  pages   = {786--789},
  year    = {2016},
  doi     = {10.1109/LPT.2015.2514078}
}

@article{Pointurier2017,
  author  = {Y. Pointurier},
  title   = {Design of Low-Margin Optical Networks},
  journal = {Journal of Optical Communications and Networking},
  volume  = {9},
  number  = {1},
  pages   = {A9--A17},
  year    = {2017},
  doi     = {10.1364/JOCN.9.0000A9}
}

@article{Mlejnek2015,
  author  = {M. Mlejnek and I. Roudas and J. D. Downie
             and N. Kaliteevskiy and K. Koreshkov},
  title   = {Coupled-Mode Theory of Multipath Interference in
             Quasi-Single-Mode Fibers},
  journal = {IEEE Photonics Journal},
  volume  = {7},
  number  = {1},
  pages   = {7100116},
  year    = {2015},
  doi     = {10.1109/JPHOT.2014.2387260}
}

@article{Downie2017,
  author  = {J. D. Downie and J. Hurley and H. {de Pedro}
             and S. Garner and J. Blaker and A. B. Zakharian
             and S. Ten and G. Mills},
  title   = {Measurements and Modeling of Multipath Interference at
             Wavelengths Below Cable Cut-Off in a {G.654} Optical Fiber
             Span},
  journal = {Optics Express},
  volume  = {25},
  number  = {8},
  pages   = {9305--9311},
  year    = {2017},
  doi     = {10.1364/OE.25.009305}
}

@article{Savory2008,
  author  = {S. J. Savory},
  title   = {Digital Filters for Coherent Optical Receivers},
  journal = {Optics Express},
  volume  = {16},
  number  = {2},
  pages   = {804--817},
  year    = {2008},
  doi     = {10.1364/OE.16.000804}
}

@inproceedings{Xie2009,
  author    = {C. Xie},
  title     = {Local Oscillator Phase Noise Induced Penalties in Optical
               Coherent Detection Systems Using Electronic Chromatic
               Dispersion Compensation},
  booktitle = {Proceedings of Optical Fiber Communication Conference (OFC)},
  year      = {2009},
  pages     = {OMT4},
  doi       = {10.1364/OFC.2009.OMT4}
}

@inproceedings{Xiong2024,
  author    = {Y. Xiong and D. Zhang and S. Gao and D. Ge
               and Y. Sun and R. Zhao and Y. Xiao and Z. Yang
               and D. Wang and H. Li and X. Duan and W. Ding and Y. Wang},
  title     = {Field-Deployed Hollow-Core Fibre Cable with
               0.11 dB/km Loss},
  booktitle = {Proceedings of European Conference on Optical
               Communications (ECOC)},
  year      = {2024},
  pages     = {Th3B.8}
}

@inproceedings{Chen2019_16384QAM,
  author    = {Xi Chen and Junho Cho and Andrew Adamiecki and Peter J. Winzer},
  title     = {{16384-QAM} Transmission at 10~{GBd} over 25-km {SSMF}
               Using Polarization-Multiplexed Probabilistic Constellation
               Shaping},
  booktitle = {Proceedings of the 45th European Conference on Optical
               Communication (ECOC)},
  year      = {2019},
  pages     = {1--4},
}

@article{Olsson2018_4096QAM200km,
  author  = {Samuel L. I. Olsson and Junho Cho and Sethumadhavan Chandrasekhar
             and Xi Chen and Peter J. Winzer and Sergejs Makovejs},
  title   = {Probabilistically Shaped {PDM} 4096-{QAM} Transmission over
             up to 200~km of Fiber Using Standard Intradyne Detection},
  journal = {Optics Express},
  volume  = {26},
  number  = {4},
  pages   = {4522--4530},
  year    = {2018},
  doi     = {10.1364/OE.26.004522}
}

@article{Chen2019_4096QAM30G,
  author  = {Xi Chen and Sethumadhavan Chandrasekhar and Junho Cho
             and Peter J. Winzer},
  title   = {Transmission of 30-{GBd} Polarization-Multiplexed
             Probabilistically Shaped 4096-{QAM} over 50.9-km {SSMF}},
  journal = {Optics Express},
  volume  = {27},
  number  = {21},
  pages   = {29916--29923},
  year    = {2019},
  doi     = {10.1364/OE.27.029916}
}

@inproceedings{Terayama2018,
  author    = {Masato Terayama and Seiji Okamoto and Keisuke Kasai
               and Masato Yoshida and Masataka Nakazawa},
  title     = {4096~{QAM} (72~{Gbit/s}) Single-Carrier Coherent Optical
               Transmission with a Potential {SE} of 15.8~b/s/Hz in
               All-Raman Amplified 160~km Fiber Link},
  booktitle = {Proceedings of Optical Fiber Communication Conference (OFC)},
  year      = {2018},
  pages     = {Th1F.2},
  doi       = {10.1364/OFC.2018.Th1F.2}
}

@article{Beppu2015_2048QAM,
  author  = {Shohei Beppu and Keisuke Kasai and Masato Yoshida and Masataka Nakazawa},
  title   = {2048~{QAM} (66~{Gbit/s}) Single-Carrier Coherent Optical
             Transmission over 150~km with a Potential {SE} of
             15.3~bit/s/Hz},
  journal = {Optics Express},
  volume  = {23},
  number  = {4},
  pages   = {4960--4969},
  year    = {2015},
  doi     = {10.1364/OE.23.004960}
}

@article{Koizumi2012_1024QAM,
  author  = {Yuki Koizumi and Kazushi Toyoda and Masato Yoshida and Masataka Nakazawa},
  title   = {1024~{QAM} (60~{Gbit/s}) Single-Carrier Coherent Optical
             Transmission over 150~km},
  journal = {Optics Express},
  volume  = {20},
  number  = {11},
  pages   = {12508--12514},
  year    = {2012},
  doi     = {10.1364/OE.20.012508}
}

@inproceedings{Ge2025_AR_HCF,
  author    = {Dawei Ge and Siyuan Liu and Qiang Qiu and Peng Li and Qiang Guo and Yiqi Li and Dong Wang and Baoluo Yan and Mingqing Zuo and Lei Zhang and Dechao Zhang and Hu Shi and Jie Luo and Han Li and Zhangyuan Chen},
  title     = {1-{Tb}/s/$\lambda$ {Transmission} over {Record} 10,714-km {AR}-{HCF}},
  booktitle = {Proceedings of the European Conference on Optical Communication (ECOC 2025)},
  year      = {2025},
  month     = {September},
  address   = {Copenhagen, Denmark},
  doi       = {10.48550/arXiv.2503.24313}
}

@inproceedings{Hong2024_HCF_terabit,
  author    = {Y. Hong and S. Almonacil and H. Mardoyan and
               C. Castineiras-Carrero and S. Osuna and J. R. Gomez
               and D. R. Knight and J. Renaudier},
  title     = {Demonstration of Beyond Terabit/s/$\lambda$ Nonlinearity-Free
               Transmission over the Hollow-Core Fibre},
  booktitle = {Proceedings of the 50th European Conference on Optical
               Communication (ECOC)},
  year      = {2024},
}

@article{Hong2025_HCF_JLT,
  author  = {Yang Hong and Sylvain Almonacil and Haik Mardoyan
             and Carina {Castineiras Carrero} and Sergio Osuna
             and Javier R. Gomez and David R. Knight and Jeremie Renaudier},
  title   = {Beyond Terabit/s/$\lambda$ Nonlinearity-Free Transmission Over
             the Hollow-Core Fiber},
  journal = {Journal of Lightwave Technology},
  volume  = {43},
  number  = {13},
  pages   = {6306--6312},
  year    = {2025},
}

@article{Galdino2017_trxnoise,
  author  = {Lidia Galdino and Daniel Semrau and Domani\v{c} Lavery
             and Gabriel Saavedra and Cristian B. Czegledi and Erik Agrell
             and Robert I. Killey and Polina Bayvel},
  title   = {On the Limits of Digital Back-Propagation in the Presence of
             Transceiver Noise},
  journal = {Optics Express},
  volume  = {25},
  number  = {4},
  pages   = {4564--4578},
  year    = {2017},
  doi     = {10.1364/OE.25.004564}
}

@misc{Sohanpal2026_launch,
  author       = {Ronit Sohanpal and Eric Sillekens and Mindaugas Jarmolovicius
                  and Robert I. Killey and Polina Bayvel},
  title        = {On the Optimum Energy-per-bit Launch Power in Coherent
                  Hollow-core Fibre Transmission Systems},
  howpublished = {arXiv:2606.17942},
  year         = {2026},
  note         = {arXiv preprint arXiv:2606.17942}
}

@inproceedings{Klaus2022_HCF,
  author    = {W. Klaus and P. J. Winzer},
  title     = {Hollow-Core Fiber Capacities with Receiver Noise Limitations},
  booktitle = {Optical Fiber Communication Conference (OFC)},
  year      = {2022},
  number    = {M2C.4},
  publisher = {Optica Publishing Group}
}

\end{document}